\documentclass[journal,onecolumn]{IEEEtran}
\usepackage{amsmath,amssymb,placeins}
\usepackage{threeparttable}
\usepackage[noadjust]{cite}
\usepackage[export]{adjustbox}
\usepackage{textcomp}
\usepackage[mathscr]{euscript}
\usepackage[compatibility=false]{caption}
\usepackage{lipsum}
\usepackage{cuted}
\usepackage{mathtools}

\usepackage{multicol}
\usepackage{array}
\usepackage{makecell}
\usepackage[utf8]{inputenc}
\usepackage{diagbox}
\usepackage{xspace}
\usepackage[symbol]{footmisc}
\usepackage{caption}
\usepackage{tabularx}

\usepackage[mathscr]{euscript}
\usepackage[usenames]{color}
\usepackage{amsfonts}
\usepackage{latexsym}
\usepackage{subfigure}
\usepackage[format=hang]{subfig}
\usepackage{caption}
\usepackage{tikz}
\usepackage{floatrow}
\usepackage{multirow}
\usepackage{epstopdf}

\newtheorem{theorem}{{{\textit{Theorem}}}}

\newtheorem{lemma}{{{\textit{Lemma}}}}
\newtheorem{corollary}{{{{\textit{Corollary}}}}}
\newcommand{\Tau}{\mathcal{T}}

\newtheorem{definition}{{{\textit{Definition}}}}

\newtheorem{remark}{{{\textit{Remark}}}}

\newtheorem{example}{{{\textit{Example}}}}

\newcommand{\Mod}[1]{\ (\mathrm{mod}\ #1)}

\begin{document}

\markboth{Journal of \LaTeX\ Class Files,~Vol.~13, No.~9, September~2014}%
{Shell \MakeLowercase{\textit{et al.}}: Bare Demo of IEEEtran.cls for Journals}

\title{{Systematic Constructions of Complementary Sets and Hadamard Matrices from Circulant Operator}}
\author{Piyush Priyanshu, Sudhan Majhi, and Subhabrata Paul}
\maketitle

\begin{abstract}
A Hadamard matrix $\mathbf{H}$ of order $n$ is a square matrix with entries $\pm 1$ satisfying $\mathbf{H}\mathbf{H}^\top = n\mathbf{I}_{n}$, where $\mathbf{I}_n$ is the identity matrix of order $n$. A circulant Hadamard matrix is a Hadamard matrix whose rows are cyclic shifts of one another. This work establishes a unified algebraic framework that treats arbitrary Hadamard matrices as flexible seeds to systematically generate Golay complementary sets (GCS), cross Z-complementary sets  (CZCS), complete complementary codes (CCC), and optimal cross-Z complementary sequence sets (CZCSS) through algebraic transformations. In this paper, a systematic framework using cyclic operators is presented. First, circulant Hadamard matrices of order 4 are utilized recursively to propose binary CZCS of arbitrary lengths, achieving a maximum ZCZ ratio of $2/3$, and binary  GCS. Significantly, this framework is generalized to establish that by employing binary or complex Hadamard matrices of any order, binary or non-binary CZCSs of arbitrary lengths can be constructed with a ZCZ ratio of $1/2$. Furthermore, to provide flexible user capacity, an alternative construction of binary GCSs of all lengths and Hadamard matrices of order $2^{a+1}10^b26^c$ ($a,b,c \geq 0$) is proposed using circulant matrices and Golay complementary pairs (GCP). These constructions are further extended to form binary CCC with parameters $(2N,2N,2N)$, where $N=2^a10^b26^c$, and $(4n,4n,4n)$ for $n \ge 1$. Additionally, optimal binary $(8n,8n,8n,4n)$-CZCSS and their complex versions with parameters {\color{red}$(2m,2m,2m,m)$ are proposed for $n, m \geq 1$.} These results provide the first generalized framework for constructing optimal CZCSS from arbitrary Hadamard seeds. Finally, a theoretical relation between Hadamard matrices and GCSs is established, and fundamental properties of circulant matrices over aperiodic correlation functions are presented. All proposed constructions are novel, and their parameters are compared with those of existing state-of-the-art methods.
\end{abstract}


\IEEEpeerreviewmaketitle

\section{Introduction}

\IEEEPARstart{T}he Hadamard conjecture states that a Hadamard matrix of order $4n$ exists for every positive integer $n \in \mathbb{N}$, a problem first proposed by Jacques Hadamard in 1893 \cite{hadamard1893resolution}. Hadamard investigated square matrices with entries $\pm 1$ characterized by the property that all rows (or columns) are pairwise orthogonal, such that $\mathbf{H} \mathbf{H}^\top=n\mathbf{I}_n$, where $\mathbf{H}$ is a square matrix of order $n$ and $\mathbf{I}_n$ is the identity matrix of the same order. Beyond the existence question, Hadamard posed the more general problem of determining the maximal determinant of matrices whose entries lie within the unit disk.

Before Hadamard's work, in 1857, Sylvester had found Hadamard matrices of orders that are powers of two \cite{sylvester1867thoughts}. Sylvester observed that if $\mathbf{H}$ is a Hadamard matrix of order $n$, then the matrix $\begin{bmatrix} \mathbf{H} & \mathbf{H} \\ \mathbf{H} & -\mathbf{H} \end{bmatrix}$ is also a Hadamard matrix of order $2n$, also known as the Sylvester construction. Sylvester's work laid the foundation for the study of Hadamard matrices, which have since found numerous applications in coding theory, signal processing, and quantum computing \cite{Hadamard2020applications}. However, Hadamard's contribution was to show the general existence conditions for Hadamard matrices.

\par The study of the construction of Hadamard matrices has attracted many researchers. In 1933, Paley provided two major theorems on the existence of Hadamard matrices, stating that if $p$ is a prime number such that $p \equiv 3 \mod 4$ and $p \equiv 1 \mod 4$, then there exists a Hadamard matrix of order $(p+1)$ and $2(p+1)$, respectively \cite{paley1933orthogonal}. In 1944, Williamson introduced matrices that later became known as Williamson-type matrices \cite{williamson1944hadamard}. In 1965, Baumert \textit{et al.} stated that a Hadamard matrix of order $12t$ exists for every Williamson-type matrix of order $4t$ \cite{BaumertHall1965}, where $t$ is a positive integer. In 1967, Goethals-Seidel proposed a strong relationship between orthogonal matrices with zero diagonal and Hadamard matrices \cite{goethals1967orthogonal}. In 1970, Cooper \textit{et al.} constructed Hadamard matrices of order $4t$, where $t \in \{1,3,5,7,\dots,19\}$ \cite{CooperWallis1972}.

\par In the same year, Turyn proposed that if there is a complex Hadamard matrix of order $2n$ and a Hadamard matrix of order $4h$, then there exists a Hadamard matrix of order $8nh$ \cite{turyn1970complex}, where {{$n,h\in \mathbb{Z}^+$}}. In 1972, Cooper \textit{et al.} provided the construction of Hadamard matrices of order $2^{t+2}q$ from T-matrices of order $2^tq$, where $q$ and $t$ are positive integers \cite{CooperWallis1972}. In 1973, Wallis constructed Hadamard matrices of order $28m$, $36m$, and $44m$ using T-matrices of order $m$ \cite{Wallis1973}, where {{$m \in \mathbb{Z}^+$}}. In 1976, W. D. Wallis established a critical connection between the existence of Hadamard matrices and Williamson-type matrices. A quadruple of symmetric circulant matrices $A$, $B$, $C$ and $D$ of order $n$, with entries $-1$ or $1$, is of Williamson type if it satisfies the following: $AA^\top + BB^\top + CC^\top + DD^\top = 4nI_n$, where $I_n$ is the identity matrix \cite{wallis1976existence}. A circulant matrix is a Toeplitz matrix where each row is a cyclic shift of the row above it \cite{toeplitz}. In 1985, Agayan-Sarukbanyan stated that if there are two Hadamard matrices of orders $4h$ and $4k$, then there exists a Hadamard matrix of order $8hk$ \cite{agaian1985hadamard}, where {{$h, k \in \mathbb{Z}^+$}}.

\par In 1989, Koukouvinous \textit{et al.} used T-matrices and a Golay complementary pair (GCP) to construct Hadamard matrices of order $2^tq$, where $q$ is the sum of the lengths of two GCPs \cite{KoukouvinosKounias1989}. In 1991, Miyamoto established the existence of Hadamard matrices of order $4q$ if there is a Hadamard matrix of order $q-1$, where $q$ is a prime power \cite{Miyamoto1991}. In 1992, Craigen \textit{et al.} showed that if there are Hadamard matrices of orders $4a$, $4b$, $4c$, $4d$, then there is a Hadamard matrix of order $16abcd$ \cite{craigen1992product}. {{More broadly, it is known that Hadamard matrices can be constructed from other types of complementary sequences, such as periodic complementary pairs (PCPs) \cite{Dok2015}.}} Many Hadamard matrices of different orders have since been found, including orders $428$ \cite{order428}, $1004$ and $2524$ \cite{okovi2013SomeNO}, $268$, $412$, $436$, and $604$ \cite{Balonin2018SymmetricHM}, and $764$ \cite{order764}. This motivated us to find new constructions of Hadamard matrices and study their structure.

{
\par A circulant matrix satisfying the Hadamard property is termed a circulant Hadamard matrix, which exists only for orders $n=1$ and $n=4$ \cite{Davis1994, morris2023proofryserscirculanthadamard,gallardo2024rysersconjecturestochasticmatrices}. In this work, the eight existing circulant Hadamard matrices of order $4$ are utilized as the foundational seeds for the systematic recursive construction of various classes of complementary sequence sets.}

\par A cross Z-complementary set (CZCS) is a set of sequences whose sum of aperiodic auto-correlation functions (AACFs) of all sequences and the sum of aperiodic cross-correlation functions (ACCFs) of adjacent sequences is zero in a specific zone called the zero-correlation zone (ZCZ) \cite{liu2020cross}. The ZCZ is defined as the ratio of the ZCZ's width to the sequence's length. CZCSs are used in spatial modulation (SM) systems in frequency-selective channels \cite{liu2020cross}. The first construction of a CZCSs of set size 4 and various lengths was provided by Huang \textit{et al.} \cite{Huang2022}, where they constructed CZCSs using concatenation. In 2022, Huang \textit{et al.} proposed CZCSs of set size 4, length $2^m$, and a ZCZ ratio of 1, where $m \geq 1$ \cite{ChenISITCZCS}. In 2023, Das \textit{et al.} used an indirect method to construct quaternary CZCSs of lengths $3L$, $7L$, and $14L$, with a ZCZ ratio less than $1/2$, where $L$ is the length of the seed sequences \cite{shibsankarCZCS}. In 2023, Kumar \textit{et al.} constructed CZCSs of length $2^{m-1}+2^{\delta}$ with a maximum ZCZ ratio of $2/3$, where $m \geq 4$ and $0 \leq \delta \leq m-1$ \cite{PraveenCZCS}. In 2025, Huang \textit{et al.} presented a flexible construction of $(2^{k+1},2^{m-k}(2^k-1)+2^v,2^{m-1})$-CZCS and $(2^{k_1+2},2^{m-1}+\sum_{\beta=1}^{k_1-1}a_{\beta}2^{\pi(m-k_1+\beta)-1}+2^{v_1},2^{m-1}+\sum_{\beta=1}^{k_1-1}a_{\beta}2^{\pi(m-k_1+\beta)-1}+2^{v_1})$-CZCS, where $m \geq 2$, $1 \leq k,k_1-1 \leq m-1$, $0 \leq v \leq m-k$, $\beta \in \mathbb{N}$ and $0 \leq v_1 \leq m-k_1$ \cite{HuangCZCS}. Although recent advancements, notably Huang \textit{et al.} in 2025 \cite{HuangCZCS}, have utilized binary decomposition of GBFs to achieve nearly universal length coverage, the existing literature relies mainly on combinatorial mapping. { While existing CZCS designs primarily utilize Boolean functions, they are typically restricted to flock sizes that are powers of two. To overcome this limitation, this work introduces a systematic framework based on arbitrary binary or complex Hadamard matrices as foundational seeds. It is demonstrated that the inherent row-orthogonality of Hadamard matrices can be algebraically transformed into CZCS. This methodology provides a direct path to constructing sequence sets with arbitrary lengths and flexible, non-power-of-two flock sizes, establishing a profound mathematical link between fundamental matrix theory and  sequence design.}

\par {{The concept of a}} Golay complementary set (GCS) was extended from the Golay complementary pair (GCP) in 1972 \cite{firstGCS1972}. A GCS is a set of sequences whose sum of the AACFs is zero at every non-zero time shift. GCSs have numerous applications, including channel estimation \cite{ISIchannel}, synchronization \cite{mimosynchronization}, and reduction of the peak-to-mean envelope power ratio (PMEPR) in orthogonal frequency division multiplexing (OFDM) systems \cite{ofdm,davis1999peak}. Due to these applications, the study of GCSs plays a vital role in wireless communication. Paterson proposed a method for the construction of GCSs using GBFs \cite{paterson2000generalized}. Initially, constructions were limited to lengths that are powers of two, until Chen proposed GCS constructions with flexible lengths using GBFs in 2016 \cite{chen2016complementary}. {{Further generalized constructions based on GCPs to achieve flexible non-power-of-two lengths have also been presented in \cite{Wang2020SPL, Pai2020EL}.}} Several constructions using GBFs, Reed-Muller codes (RMC), generalized Reed-Muller codes (GRM) and extended Boolean functions have been proposed in \cite{wang2017method,chen2017new,chen2018novel,adhikary2019new,chen2021GCS}. In \cite{ma2017new}, the authors introduced para-unitary (PU) matrices, where each element has a unit magnitude, as a new method to construct GCSs. A Hadamard matrix is a special case of a PU matrix. Wang \textit{et al.} made significant progress in the construction of GCSs using PU matrices \cite{wang2021new,wang2023constructions}. However, no construction covering all lengths existed until Roy \textit{et al.} proposed a binary GCS construction covering all lengths, {{albeit with flock sizes restricted to powers of 2}} \cite{abhishekGCS}. All known binary GCS constructions to date have flock sizes limited to powers of 2. This limitation motivated the development of binary GCS constructions for all lengths with a flock size equal to $2N$, where $N = 2^a10^b26^c$, $a,b,c \geq 0$, {{a goal that has not yet been achieved.}}

{
\par In 1988, Suehiro and Hatori extended the concept of GCSs to establish complete complementary codes (CCC) \cite{concccunitarymatrices}. A CCC is characterized by the parameters $(N, N, L)$, where $N$ represents the code size and $L$ sequence length, respectively, and is widely utilized in coding, signal processing, and wireless communications \cite{appshib28,appshib31,appshib33,appshib34}. Subsequent research has yielded diverse construction methodologies, including foundational work based on Reed-Muller codes for power-of-two sizes \cite{Rathinakumar2008}, as well as algebraic approaches utilizing GBFs and permutation polynomials \cite{concccBooleanfunction, sarkar2021multivariable, wang2023constructions}. More recently, general frameworks based on PU matrices have been developed to generate CCCs with flexible parameters \cite{wang2021new, concccPUmatrices, shibsankarvarious}. Although existing methodologies for constructing $(N,N,N)$-CCCs offer significant flexibility, they often involve intricate generation procedures or results in non-binary sequence alphabets. For instance, general PU-based frameworks typically utilize normalized Hadamard matrices to satisfy specific unitary conditions. In contrast, this paper establishes a simplified algebraic framework that utilizes Hadamard matrices directly. By bypassing the requirement for normalization, the proposed approach provides a straightforward path to constructing strictly binary CCCs with flexible, non-power-of-two flock sizes.} 

\par In 2024, Kumar \textit{et al.} proposed an extension of CZCSs and a generalization called symmetrical Z-complementary code sets (SZCCS) \cite{praveenCZCCS}. The authors proposed a direct construction of $(2^{n+1},2^{n+1},2^{m-1}+2,2^{\pi(m-3)})$-CZCSS, where $n \geq 0$ and $m \geq 4$. In the same year, Huang \textit{et al.} proposed three major constructions of CZCSSs and named them Enhanced Cross Z-Complementary Sets (E-CZCSs). These constructions include $(M,N,2L,Z)$-CZCSS using a $(M,N,L,Z+1)$-ZCCS, $(M,N,2L,L)$-CZCSS, and $(2^k,2^v,2^m,2^{\pi_1(1)-1})$-CZCSS. The construction is optimal when $\pi_1(1)=m-k+v$, where {{$m,k,v \in \mathbb{Z}^+$}} and $v \leq k$ \cite{EnhancedZLiu}. To date, no optimal construction of CZCSSs of lengths not in the power of two is available.

\par In this paper, we propose new constructions of Hadamard matrices, circulant Hadamard matrices, CZCSs, GCSs, CCCs, and CZCSSs by using linear operators and circulant matrices as follows:

{\begin{itemize}

    \item A recursive framework utilizing circulant Hadamard matrices of order 4 as foundational seeds is proposed to construct binary GCSs and CZCSs of arbitrary length. The proposed CZCSs achieve a maximum ZCZ ratio of $2/3$, providing a systematic construction path for even-phase sequences.
    
    \item A generalized framework is established to prove that any binary or complex Hadamard matrix can be algebraically transformed into an arbitrary length CZCS with a ZCZ ratio of $1/2$. Specifically, we show that any Hadamard matrix $\mathbf{H}_{4n}$ ($n \ge 1$) can be utilized to construct a recursive Hadamard matrix $\mathbf{H}_{8n} = \mathbf{H}_2 \otimes \mathbf{H}_{4n}$, whose truncated rows form a binary $(8n, 8n-k, 4n-k)$-CZCS for $0 \leq k \leq 4n-1$.

    \item A second construction framework utilizing circulant matrices seeded by GCPs is introduced to generate GCSs with flexible, non-power-of-two flock sizes of the form $2N$, where $N=2^a10^b26^c$ for $a,b,c \ge 0$. This systematic approach overcomes the power-of-two flock size limitations inherent in much of the prior art.

    \item The GCSs generated from both proposed frameworks are demonstrated to be inherently Hadamard matrices. These are further utilized to construct two new families of binary complete complementary codes (CCC): a flexible $(2N,2N,2N)$-CCC and a general $(4n,4n,4n)$-CCC for any integer $n \ge 1$. {\color{red}We also generalize the result for unimodular $(m,m,m)$-CCC, where $m\geq1$. }

     \item A new extension of the generalized construction of CZCS is proposed to generate optimal binary $(8n,8n,8n,4n)$-CZCSSs and their complex counterparts with parameters {\color{red}$(2m,2m,2m,m)$ for $n,m \ge 1$. }This constitutes the first universal mechanism capable of converting arbitrary Hadamard seeds into optimal cross-Z complementary sequence sets, achieving the theoretical ZCZ bound.

    \item The theoretical relationship between GCSs and Hadamard matrices is formalized, and the fundamental properties of circulant matrices over aperiodic correlation functions are established to support the logical rigor of the proposed theorems.
     \item A unified transformation pathway is established, demonstrating that GCS, CZCS, CCC, and the optimal CZCSS share a common mathematical origin. We provide a systematic diagram illustrating that these diverse complementary structures can be reached from a single Hadamard seed using three fundamental algebraic operations: expansion, truncation, and row-wise transformation.
\end{itemize}}


\par {The rest of this paper is organized as follows. Section II reviews the essential definitions and preliminaries. Section III establishes several key lemmas on the correlation properties of circulant matrices. The constructions for complementary sets and Hadamard matrices are presented in Section IV and compared with the state-of-the-art in Section V. Finally, Section VI concludes the paper and discusses open problems.}

\section{Preliminaries}
This section introduces essential definitions, notations, and theorems used in the constructions presented later in the paper.
\subsection{Definitions}
\begin{definition}
{{Let \(\mathbf{a} = (a_0, a_1, \dots, a_{L-1})\) and \(\mathbf{b} = (b_0, b_1, \dots, b_{L-1})\) be sequences of length \(L\).}} The aperiodic cross-correlation function (ACCF) is defined as
\begin{equation}\label{eqn1}
\mathcal{C}(\mathbf{a}, \mathbf{b})(\lambda) =
\begin{cases}
{{\sum\limits_{i=0}^{L-1-\lambda} a_i b^*_{i+\lambda},}} & 0 \leq \lambda \leq L-1, \\
{{\sum\limits_{i=0}^{L-1+\lambda} a_{i-\lambda} b^*_{i},}} & -L+1 \leq \lambda \leq -1, \\
0, & |\lambda| \geq L,
\end{cases}
\end{equation}
where $b^*$ represents the complex conjugate of $b$. When \(\mathbf{a} = \mathbf{b}\), ACCF becomes the aperiodic autocorrelation function (AACF), denoted \(\mathcal{A}(\mathbf{a})(\lambda)\).
\end{definition}
\begin{definition}
Let $\mathbf{G}$ be an $M \times L$ matrix whose rows are the sequences $\mathbf{a}_0, \mathbf{a}_1, \dots, \mathbf{a}_{M-1}$. The set of these row sequences is called a Golay complementary set (GCS) if
\begin{equation}
\sum\limits_{j=0}^{M-1} \mathcal{A}(\mathbf{a}_j)(\lambda) = 0, \quad \forall \lambda \neq 0.
\end{equation}
\end{definition}
When \(M = 2\), the set of rows is referred to as a Golay complementary pair (GCP).

\begin{definition}
Let \((\mathbf{a}, \mathbf{b})\) and \((\mathbf{c}, \mathbf{d})\) be two GCPs of length \(L\). The pair \((\mathbf{c}, \mathbf{d})\) is a complementary mate of \((\mathbf{a}, \mathbf{b})\) if
\begin{equation}
\mathcal{C}(\mathbf{a}, \mathbf{c})(\lambda) + \mathcal{C}(\mathbf{b}, \mathbf{d})(\lambda) = 0, \quad \forall \lambda \neq 0.
\end{equation}
\end{definition}

\begin{definition}
Let $\mathbf{G}$ be an $M \times L$ matrix whose rows are the sequences $\mathbf{a}_0, \mathbf{a}_1, \dots, \mathbf{a}_{M-1}$. The set of these row sequences is called a \((M, L, Z)\)-cross Z-complementary set (CZCS) if it satisfies the following conditions:
\begin{equation}
\begin{aligned}
\sum\limits_{j=0}^{M-1} \mathcal{A}(\mathbf{a}_j)(\lambda) &= 0, \quad \forall |\lambda| \in \Tau_1 \cup \Tau_2, \\
\sum\limits_{j=0}^{M-1} \mathcal{C}(\mathbf{a}_j, \mathbf{a}_{(j+1) \mod M})(\lambda) &= 0, \quad \forall |\lambda| \in \Tau_2,
\end{aligned}
\end{equation}
\end{definition}
where \(\Tau_1 = \{1, 2, \dots, Z\}\) and \(\Tau_2 = \{L-Z, L-Z+1, \dots, L-1\}\). When \(M = 2\), the set reduces to a cross Z-complementary pair (CZCP). The ratio \(Z / L\) is called the ZCZ ratio.

{
{
\begin{definition}[\cite{Davis1994}]
A left cyclic shift operator \(T^k: \mathbb{C}^n \rightarrow \mathbb{C}^n\) is defined for a vector \(\mathbf{v} = (v_0, v_1, \dots, v_{n-1})\). For \(1 \leq k \leq n\), the \(k\)-th vector of this operator is given by:
\begin{equation}
T^k(\mathbf{v}) = (v_{k-1}, v_{k-2}, \dots, v_0, v_{n-1}, v_{n-2}, \dots, v_k).
\end{equation}
Examples of this operator for the first two applications are:
\begin{itemize}
    \item $T^1(\mathbf{v}) = (v_0, v_{n-1}, v_{n-2}, \dots, v_1)$,
    \item $T^2(\mathbf{v}) = (v_1, v_0, v_{n-1}, \dots, v_2)$,
\end{itemize}
and for the final application, $T^n(\mathbf{v}) = (v_{n-1}, v_{n-2}, \dots, v_0)$. This operator is utilized to systematically generate the row sequences of the circulant structures proposed in this work.
\end{definition}
}


{
\begin{definition}
For consistency with the notation used in the proofs, a second operator is defined, the right cyclic shift operator \(T_1^k: \mathbb{C}^n \rightarrow \mathbb{C}^n\). Applying this operator $k$ times to a vector $\mathbf{v}=(v_0, v_1, \dots, v_{n-1})\in \mathbb{C}^n$ results in the column vector $(T_1^k(\mathbf{v}))^\top$, where the $j$-th entry is given by $v_{(j-k) \pmod n}$ for $0 \le j \le n-1$. Examples of this operation include:
\begin{itemize}
    \item $(T_1^0(\mathbf{v}))^\top = (v_0, v_1, \dots, v_{n-1})^\top$,
    \item $(T_1^1(\mathbf{v}))^\top = (v_{n-1}, v_0, v_1, \dots, v_{n-2})^\top$,
    \item $(T_1^2(\mathbf{v}))^\top = (v_{n-2}, v_{n-1}, v_0, \dots, v_{n-3})^\top$.
\end{itemize}
For $k \ge 1$, the general term is expressed as:
\begin{equation}
(T_1^k(\mathbf{v}))^\top = (v_{n-k}, v_{n-k+1}, \dots, v_{n-1}, v_0, v_1, \dots, v_{n-k-1})^\top.
\end{equation}
\end{definition}
}

\begin{remark}
In this framework, $T$ serves as the primary operator for constructing the sequence sets, while $T_1$ is specifically defined to represent the columns of the circulant matrix, facilitating the derivation of correlation properties in the Appendices.
\end{remark}

\begin{definition}
A circulant matrix of order $n$, denoted by \(\text{Cir}(\mathbf{a})\), is constructed from a seed sequence $\mathbf{a}$. The rows of the matrix are generated by the left shift operator $T$, while the columns are represented by the right shift operator $T_1$:
\begin{equation}
\text{Cir}(\mathbf{a}) =
\begin{bmatrix}
T^{1}(\mathbf{a}) \\
T^{2}(\mathbf{a}) \\
\vdots \\
T^{n}(\mathbf{a})
\end{bmatrix} =
\begin{bmatrix}
(T_1^0(\mathbf{a}))^\top & (T_1^1(\mathbf{a}))^\top & \cdots & (T_1^{n-1}(\mathbf{a}))^\top
\end{bmatrix}.
\end{equation}
\end{definition}

\begin{example}
Let \(\mathbf{a} = (1, -1, -1, 1)\). The circulant matrix \(\text{Cir}(\mathbf{a})\) is given by:
\begin{equation*}
\text{Cir}(\mathbf{a}) =
\begin{bmatrix}
1 & 1 & -1 & -1 \\
-1 & 1 & 1 & -1 \\
-1 & -1 & 1 & 1 \\
1 & -1 & -1 & 1
\end{bmatrix}.
\end{equation*}
\end{example}
}

\begin{definition}
A matrix \(H\) of size \(n \times n\), whose entries are either \(1\) or \(-1\), is called a Hadamard matrix if it satisfies the following:
\begin{equation}
H H^\top = n I_{n},
\end{equation}
where \(I_n\) is the identity matrix of order \(n\). It is conjectured that Hadamard matrices exist for orders $1, 2$, and all multiples of $4$, {\color{red}whereas complex Hadamard matrices exist for every positive integer \cite{ComplexHadamard}.}
\end{definition}

\begin{example}
Consider the matrix:
\begin{equation*}
H = \begin{bmatrix}
1 & 1 & 1 & 1 \\
1 & -1 & 1 & -1 \\
1 & 1 & -1 & -1 \\
1 & -1 & -1 & 1
\end{bmatrix}
\end{equation*}
Then:
\begin{equation*}
H H^\top = 4 I_4,
\end{equation*}
Verify that \(H\) is a Hadamard matrix of order 4.
\end{example}

\begin{definition}
Let \(S = \{ S^0, S^1, \dots, S^{N-1} \}\), where each \(S^p = \{ \mathbf{a}^p_0, \mathbf{a}^p_1, \dots, \mathbf{a}^p_{M-1} \}\) is a set of \(M\) sequences of length \(L\). The collection \(S\) is called a mutually orthogonal Golay complementary set (MOGCS), denoted \((N, M, L)\)-MOGCS, if it satisfies:
\begin{equation}
\mathcal{C}(S^p, S^{p'})(\lambda) = \sum\limits_{i=0}^{M-1} \mathcal{C}(\mathbf{a}^p_i, (\mathbf{a}^{p'}_i)^*)(\lambda) =
\begin{cases}
ML, & \lambda = 0,~ p = p'; \\
0, & \text{otherwise}.
\end{cases}
\end{equation}
When \(N = M\), the MOGCS becomes a complete complementary code (CCC), denoted \((N, N, L)\)-CCC.
\end{definition}

\begin{definition}
 Let $S=\{S^0,S^1,\dots,S^{N-1}\}$ be a collection of $N$ sequence sets, where each $S^p$ contains $M$ sequences of length $L$, i.e., $S^p=\{\mathbf{a}^p_0, \mathbf{a}^p_1,\dots,\mathbf{a}^p_{M-1}\}$, for $0\leq p \leq N-1$. The collection $S$ is called a $(K,M,L,Z)$ cross-Z complementary sequence set (CZCSS) if each set $S^p$ satisfies the following four conditions:
  \begin{align}
        \sum_{j=0}^{M-1}{\mathcal{A}\left(\mathbf{a}^p_j\right)(\lambda)} &= 0, && \forall~ |\lambda| \in \left(\Tau_1 \cup \Tau_2\right) \cap \Tau, \\
        \sum_{j=0}^{M-1}{\mathcal{C}\left(\mathbf{a}^p_j,\mathbf{a}^p_{(j+1)\pmod M}\right)(\lambda)} &= 0, && \forall~ |\lambda| \in \Tau_2, \\
        \sum_{j=0}^{M-1}{\mathcal{C}\left(\mathbf{a}^p_j,\mathbf{a}^{p'}_j\right)(\lambda)} &= 0, && \forall~ |\lambda| \in \{0\} \cup \Tau_1 \cup \Tau_2, \label{equation13} \\
        \sum_{j=0}^{M-1}{\mathcal{C}\left(\mathbf{a}^p_j,\mathbf{a}^{p'}_{(j+1)\pmod M}\right)(\lambda)} &= 0, && \forall~ |\lambda| \in \Tau_2, \label{eqa14}
  \end{align}
    where $\Tau_1=\{1,2,\dots,Z\}$, $\Tau_2=\{L-Z,L-Z+1,\dots,L-1\}$, and $\Tau=\{1,2,\dots,L-1\}$. Optimality is achieved when $Z=NL/2M$ for binary sequences \cite{EnhancedZLiu}.
\end{definition}

\subsection{Generalized Boolean Function}

A Generalized Boolean Function (GBF) is a mapping \( f: \mathbb{Z}_2^m \to \mathbb{Z}_q \), where \( m \geq 1 \), \( q \in 2\mathbb{Z}^+ \), and \( x_i \in \{0,1\} \) for all \( 1 \leq i \leq m \). The domain consists of all binary vectors of length \( m \), and the co-domain is the ring of integers modulo \( q \).

We define \( 2^m \) monomials of degree \( r \) as all possible products of up to \( r \) distinct variables from \( \{x_1, x_2, \dots, x_m\} \). Examples include
\begin{itemize}
    \item Degree 0: \( 1 \)
    \item Degree 1: \( x_1, x_2, \dots, x_m \)
    \item Degree 2: \( x_1x_2, x_2x_3, \dots, x_{m-1}x_m \)
    \item Degree \( m \): \( x_1x_2\dots x_m \)
\end{itemize}

It is established in \cite{Boolean} that any GBF \( f \) can be uniquely represented as a linear combination of these monomials. The sequence \( \mathbf{f} = (f_0, f_1, \dots, f_{2^m-1}) \) is generated by evaluating \( f \) at all binary inputs, where each \( f_I = f(i_1, i_2, \dots, i_m) \) and \( (i_1, i_2, \dots, i_m) \) is the binary representation of the integer \( I \in [0, 2^m - 1] \), calculated by:
\begin{equation}
I = \sum_{k=1}^{m} i_k 2^{k-1}.
\end{equation}

The associated complex-valued sequence is \( \psi(\mathbf{f}) = (\xi^{f_0}, \xi^{f_1}, \dots, \xi^{f_{2^m-1}}) \), {where \( \xi = e^{2\pi \sqrt {-1} / q} \).}

\begin{example}
Let \( m = 3 \) and \( q = 2 \). A Boolean function \( f: \mathbb{Z}_2^3 \to \mathbb{Z}_2 \) yields the following sequence:
\begin{equation*}
\begin{aligned}
\mathbf{f} &= \big(f(0, 0, 0), f(1, 0, 0), f(0, 1, 0), f(1, 1, 0), 
 f(0, 0, 1), f(1, 0, 1), f(0, 1, 1), f(1, 1, 1)\big).
\end{aligned}
\end{equation*}
Its complex representation is as follows 
\begin{equation*}
\psi(\mathbf{f}) = \left((-1)^{f_0}, (-1)^{f_1}, \dots, (-1)^{f_7}\right),
\end{equation*}
where \( \xi = e^{2\pi \sqrt{-1} / 2} = -1 \).
\end{example}

\subsection{Truncation}

Let \( \mathbf{A} = [a_{ij}]_{m \times n} \) be a matrix. We define its truncation \( \mathbf{A}^k \) by removing the last columns \( k \):
\begin{equation*}
\mathbf{A}^k = [a_{ij}]_{m \times (n-k)}, \quad 1 \leq i \leq m,\ 1 \leq j \leq n-k.
\end{equation*}

For a sequence \( \mathbf{a} \) of length \( L \), the truncated sequence is denoted \( \mathbf{a}^{L-k} \), indicating that the last \( k \) elements are removed \cite{truncation}.

\subsection{Lemmas}

\begin{lemma}[\cite{davis3}]
Let \( \pi \) be a permutation of \( \{1, 2, \dots, m\} \), where \( m \geq 1 \). Define a GBF:
\begin{equation}
f(x_1, x_2, \dots, x_m) = 2^{h-1}\sum_{i=1}^{m-1} x_{\pi(i)}x_{\pi(i+1)} + \sum_{k=1}^{m} c_k x_k,
\end{equation}
where \( h \geq 1 \), and each \( c_k \in \mathbb{Z}_2 \). Then the sequences \( \mathbf{a} = \psi(f + \theta) \) and \( \mathbf{b} = \psi(f + 2^{h-1}x_{\pi(1)} + \theta') \), for arbitrary \( \theta, \theta' \in \mathbb{Z}_2 \), form a GCP of length \( 2^m \).
\end{lemma}

\begin{lemma}[\cite{TURYN1974313}]\label{lemmapre2}
Let \( (\mathbf{a}, \mathbf{b}) \) and \( (\mathbf{c}, \mathbf{d}) \) be GCPs of lengths \( m \) and \( n \), respectively. Then the pair:
\begin{equation*}
\mathbf{e} = \mathbf{a} \otimes \left(\frac{\mathbf{c} + \mathbf{d}}{2}\right) - \mathbf{b}^* \otimes \left(\frac{\mathbf{c} - \mathbf{d}}{2}\right),
\end{equation*}
and
\begin{equation*}
\mathbf{f} = \mathbf{b} \otimes \left(\frac{\mathbf{c} + \mathbf{d}}{2}\right) + \mathbf{a}^* \otimes \left(\frac{\mathbf{c} - \mathbf{d}}{2}\right)
\end{equation*}
form a GCP of length \( mn \). Here, \( \otimes \) denotes the Kronecker product, and \( \mathbf{a}^* \) denotes the complex conjugate.
\end{lemma}

\begin{lemma}[\cite{lemma14}]\label{lemmapre3}
Let \( (\mathbf{a}, \mathbf{b}) \) be a GCP of length \( L \). Then \( (\mathbf{c}, \mathbf{d}) = (\overleftarrow{\mathbf{b}^*}, -\overleftarrow{\mathbf{a}^*}) \) is a complementary mate of \( (\mathbf{a}, \mathbf{b}) \).
\end{lemma}

\section{Proposed Lemmas}
In this section, we propose several lemmas that play a crucial role in proving the subsequent theorems.

{
\begin{lemma}\label{Lemma4}
    Let $(T_1^i(\mathbf{a}))^\top$ and $(T_1^j(\mathbf{b}))^\top$ denote the column vectors $i$ -th and $j$ -th of the circulant matrices $\text{Cir}(\mathbf{a})$ and $\text{Cir}(\mathbf{b})$, respectively. For $j \leq i$, the inner product is given by:
    \begin{equation}
        \langle (T_1^{i}(\mathbf{a}))^\top, (T_1^{j}(\mathbf{b}))^\top \rangle = T_1^{i}(\mathbf{a}) (T_1^{j}(\mathbf{b}))^\dagger = \mathcal{C}\left(\mathbf{b}, \mathbf{a}\right)\left(L-k'\right) + \mathcal{C}\left(\mathbf{a}, \mathbf{b}\right)\left(k'\right),
    \end{equation}
    where $k' = i - j$ and $L$ is the length of the sequences and $\dagger$ denotes the conjugate transpose,.
\end{lemma}
\begin{IEEEproof}
    See Appendix A.
\end{IEEEproof}
\begin{lemma}\label{Lemma5}
    For $i \leq j$, the inner product of the column vectors $(T_1^i(\mathbf{a}))^\top$ and $(T_1^j(\mathbf{b}))^\top$ is:
    \begin{equation}
        \langle (T_1^{i}(\mathbf{a}))^\top, (T_1^{j}(\mathbf{b}))^\top \rangle = T_1^{i}(\mathbf{a}) (T_1^{j}(\mathbf{b}))^\dagger = \mathcal{C}\left(\mathbf{b}, \mathbf{a}\right)\left(k'\right) + \mathcal{C}\left(\mathbf{a}, \mathbf{b}\right)\left(L-k'\right),
    \end{equation}
    where $k' = j - i$.
\end{lemma}
\begin{IEEEproof}
  The proof is similar to that of Lemma \ref{Lemma4}.
\end{IEEEproof}
\begin{lemma}\label{Lemma6}
    The inner product of two cyclic shifts of the same row sequence for $i \neq j$ is
    \begin{equation}
    \langle T^{i}(\mathbf{a}), T^{j}(\mathbf{a}) \rangle = T^{i}(\mathbf{a}) (T^{j}(\mathbf{a}))^\dagger = \mathcal{A}(\mathbf{a})(k') + \mathcal{A}(\mathbf{a})(L-k'), 
    \end{equation}
    where $k' = |j-i|$. Due to circulant symmetry, the same relationship holds for the column vectors:
    \begin{equation}
        \langle (T_1^{i}(\mathbf{a}))^\top, (T_1^{j}(\mathbf{a}))^\top \rangle = T_1^{i}(\mathbf{a}) (T_1^{j}(\mathbf{a}))^\dagger = \mathcal{A}\left(\mathbf{a}\right)\left(k'\right) + \mathcal{A}\left(\mathbf{a}\right)\left(L-k'\right).
    \end{equation}
\end{lemma}
\begin{IEEEproof}
  The proof is similar to that of Lemma \ref{Lemma4}.
\end{IEEEproof}
}

\begin{lemma}\label{Lemma7}
    Let $\mathbf{A}=Cir(\mathbf{a})$ be a circulant matrix of order $n$ corresponding to a sequence $\mathbf{a}$. The sum of AACFs of the rows of the truncated matrix $\mathbf{A}^k$ is given by
    \begin{equation}\label{Lemma722}
    \begin{aligned}
        \sum_{i=1}^{n}{\mathcal{A}\left(T^i\left(\mathbf{a}\right)\right)\left(\lambda \right)}&= \left(n-\lambda-k\right)\left(\mathcal{A}\left(\mathbf{a}\right)\left(\lambda\right)   ~+\mathcal{A}\left(\mathbf{a}\right)\left(n-\lambda\right)\right) , ~~~ 0\leq \lambda \leq n-1-k,
    \end{aligned}
    \end{equation}
    where $0\leq k\leq n-1$.
\end{lemma}
\begin{IEEEproof}
    See Appendix B.
\end{IEEEproof}
\begin{lemma}\label{Lemma8}
    Let $\mathbf{a}$ and $\mathbf{b}$ be sequences of length $n$, and consider the matrix $\mathbf{Z}=[Cir(\mathbf{a}),Cir(\mathbf{b})]$. The sum of AACFs of the rows of the truncated matrix $\mathbf{Z}^k$, denoted $R_i$, where $1\leq i \leq n$, is given by 
    \begin{equation}\label{Theorem423}
        \begin{aligned}
        \sum_{i=0}^{n}\mathcal{A}\left(R_i\right)\left(\lambda\right)= &\begin{cases}
                (n-\lambda)(\mathcal{A}(\mathbf{a})(\lambda)+\mathcal{A}(\mathbf{a})(n-\lambda))+(n-\lambda-k)(\mathcal{A}(\mathbf{b})(\lambda)+\mathcal{A}(\mathbf{b})(n-\lambda))\\+\lambda(\mathcal{C}(\mathbf{a},\mathbf{b})(\lambda)+\mathcal{C}(\mathbf{a},\mathbf{b})(n-\lambda), ~~ \lambda \leq n-1,&\\
                (\lambda-k)(\mathcal{A}(\mathbf{a})(\lambda\mod n)+\mathcal{A}(\mathbf{a})(n-\lambda \mod n)), ~ \lambda\geq n,
            \end{cases}
        \end{aligned}
    \end{equation}
    for $0\leq k \leq n-1$, and by
    \begin{equation}\label{Theorem424}
        \begin{aligned}
      \sum_{i=0}^{n}\mathcal{A}\left(R_i\right)\left(\lambda\right)= 
               (\lambda-k\mod n)(\mathcal{A}(\mathbf{a})(\lambda)+\mathcal{A}(\mathbf{a})(n-\lambda)),
               ~\forall \lambda
        \end{aligned}
    \end{equation}
    for $n\leq k \leq 2n-2$.
    \begin{IEEEproof}
    See Appendix C.
\end{IEEEproof}
\end{lemma}
                

\begin{lemma}\label{Lemma10}
    Let $\mathbf{D} = \begin{bmatrix} \text{Cir}(\mathbf{a}) \\ \text{Cir}(\mathbf{c}) \end{bmatrix}$ be a $2L \times L$ matrix. For row indices satisfying $1 \leq i \leq L < j \leq 2L$, the inner product of the corresponding rows is:
    \begin{equation}
        \langle T^{i}(\mathbf{a}), T^{j}(\mathbf{c}) \rangle = T^{i}(\mathbf{a}) \cdot (T^{j}(\mathbf{c}))^* = \mathcal{C}\left(\mathbf{a}, \mathbf{c}\right)\left(k'\right) + \mathcal{C}\left(\mathbf{c}, \mathbf{a}\right)\left(L-k'\right),
    \end{equation}
    where $k' = (j - i) \pmod L$.
\end{lemma}
\begin{IEEEproof}
     The proof is similar to that of {{Lemma \ref{Lemma4}}}.
\end{IEEEproof}
\section{Proposed Constructions}


{
\subsection{Proposed Construction of CZCSs}
In this sub-section, a recursive framework for the construction of CZCSs and GCSs is established. The construction is initiated using circulant Hadamard matrices of order $4$, the existence of which is a classical result in combinatorial design \cite{toeplitz}. These matrices serve as foundational seeds for the power-of-two expansions developed in this framework.

\begin{theorem}\label{Theorem8}
    Let $\mathbf{E}_4$ be a circulant Hadamard matrix of order $4$. Let be the Hadamard matrix of order 2. A sequence of matrices is defined recursively as
    \begin{equation}
         \mathbf{E}_{2^{n+2}} = \mathbf{H}_2 \otimes \mathbf{E}_{2^{n+1}}, \quad \text{for } n \geq 1,
    \end{equation}
    where $\otimes$ denotes the Kronecker product, and the recursion starts with $\mathbf{E}_{2^{1+1}} = \mathbf{E}_4$.
    
       Then, the set of row sequences from the truncated matrix $\mathbf{E}^k_{2^{n+2}}$ forms a $\left(2^{n+2}, 2^{n+2}-k, Z\right)$-CZCS, where $Z = 2^{n+1}-\left(k-2^n\right)\lfloor\frac{k}{2^n}\rfloor$, $n\geq1$, and $0\leq k \leq 2^{n+1}-1$.
\end{theorem}

\begin{IEEEproof}
    See Appendix D.
\end{IEEEproof}
\begin{theorem}\label{Theorem9_GeneralizedCZCS}
    Let $\mathbf{H}_{4n}$ be any Hadamard matrix of order $4n$, where $n \geq 1$. Define the recursive Hadamard matrix $\mathbf{H}_{8n} = \mathbf{H}_2 \otimes \mathbf{H}_{4n}$. Then, the set of row sequences from the truncated matrix $\mathbf{H}^k_{8n}$ forms a binary $(8n, 8n-k, 4n-k)$-CZCS for $0 \leq k \leq 4n-1$.
\end{theorem}
\begin{IEEEproof}
    See Appendix E.
\end{IEEEproof}
{\color{red}\begin{remark}\label{RemarkCZCS}
    The framework presented in \textit{Theorem} \ref{Theorem9_GeneralizedCZCS} can be extended further to the complex domain. If $\mathbf{H}_{m}$ is a complex Hadamard matrix of order $m$, where $m \geq 1$, the rows of the truncated matrix constructed as $\mathbf{H}^k_{2m} = (\mathbf{H}_2 \otimes \mathbf{H}_{m})^k$ form a non-binary $(2m, 2m-k, m-k)$-CZCS for $0 \leq k \leq m-1$.
\end{remark}}
}
{
\begin{example}\label{Example_CZCS_Recursive}
    Consider the recursive construction of a CZCS for $n=1$, which yields a set of size $2^{1+2}=8$. Let the seed circulant Hadamard matrix $\mathbf{E}_4$ be defined as:
    \begin{equation}\label{explicit_E4_seed}
        \mathbf{E}_4 = \begin{bmatrix}
            -1 & 1 & 1 & 1 \\
             1 & -1 & 1 & 1 \\
             1 & 1 & -1 & 1 \\
             1 & 1 & 1 & -1
        \end{bmatrix}.
    \end{equation}
    Using the order-2 Hadamard matrix $\mathbf{H}_2 = \begin{bsmallmatrix} 1 & 1 \\ 1 & -1 \end{bsmallmatrix}$, the order-8 Hadamard matrix is constructed via the Kronecker product $\mathbf{E}_8 = \mathbf{H}_2 \otimes \mathbf{E}_4$:
    \begin{equation*}
       \mathbf{E}_8 = \begin{bmatrix} \mathbf{E}_4 & \mathbf{E}_4 \\ \mathbf{E}_4 & -\mathbf{E}_4 \end{bmatrix} = 
        \begin{bmatrix}
            -1 &  1 &  1 &  1 & -1 &  1 &  1 &  1 \\
             1 & -1 &  1 &  1 &  1 & -1 &  1 &  1 \\
             1 &  1 & -1 &  1 &  1 &  1 & -1 &  1 \\
             1 &  1 &  1 & -1 &  1 &  1 &  1 & -1 \\
            -1 &  1 &  1 &  1 &  1 & -1 & -1 & -1 \\
             1 & -1 &  1 &  1 & -1 &  1 & -1 & -1 \\
             1 &  1 & -1 &  1 & -1 & -1 &  1 & -1 \\
             1 &  1 &  1 & -1 & -1 & -1 & -1 &  1
        \end{bmatrix}.
    \end{equation*}
    According to \textit{Theorem} \ref{Theorem8}, the set of row sequences from the truncated matrix $\mathbf{E}^k_8$ forms an $(8, 8-k, Z)$-CZCS for $0 \leq k \leq 3$, where the zero correlation zone is $Z = 4 - (k-2)\lfloor \frac{k}{2} \rfloor$. Specifically, for $k=0$, the matrix $\mathbf{E}_8$ is a Hadamard matrix of order 8, and its rows satisfy the conditions for an $(8, 8, 4)$-CZCS.
\end{example}
}
\begin{corollary}{\label{Corollary6}}
   Furthermore, the set of row sequences from the truncated matrix $\mathbf{E}^k_{2^{n+2}}$ also forms a $(2^{n+2},2^{n+2}-k)$-GCS, where $n\geq1$ and $0\leq k \leq 2^{n+1}-2$.
\end{corollary}
\begin{IEEEproof}
    This proof can be seen in the proof of \textit{Theorem} \ref{Theorem8}.
\end{IEEEproof}
\begin{remark}\label{Remark1}
    {The recursive construction defined in \textit{Theorem} \ref{Theorem8}, $\mathbf{E}_{2^{n+2}} = \mathbf{H}_2 \otimes \mathbf{E}_{2^{n+1}}$, is the standard Sylvester construction. A well-known property of the Kronecker product is that if $\mathbf{A}$ and $\mathbf{B}$ are Hadamard matrices, then $\mathbf{A} \otimes \mathbf{B}$ is also a Hadamard matrix. Since construction starts with a Hadamard matrix $\mathbf{E}_4$ and recursively applies this product to $\mathbf{H}_2$, the resulting matrix $\mathbf{E}_{2^{n+2}}$ is inherently a Hadamard matrix of order $2^{n+2}$ for all $n \geq 1$.}
\end{remark}

\begin{remark}\label{Remark2}
  From {{\textit{Theorem} \ref{Theorem8}}}, {{Corollary \ref{Corollary6}}}, and {{Remark \ref{Remark1}}}, we find that there exist $\left(2^{n+2},2^{n+2},2^{n+1}\right)$-CZCSs and $(2^{n+2},2^{n+2})$-GCSs that are also Hadamard matrices of order $2^{n+2}$, for $n\geq1$. The relationship is depicted in Fig. \ref{fig:venn_intersection}.
\end{remark}

\begin{table*}[ht!]
\begin{threeparttable}[t]
\centering
\caption{Representation of each matrix $R_i\odot \mathbf{G}$, for all $1\leq i \leq 8$, generated from Example \ref{Example8}.}
\resizebox{\textwidth}{!}{
\begin{tabular}{|l|l|l|l|}
\hline \hspace{2.5cm}$R_1\odot \mathbf{G}$ & \hspace{2.5cm}$R_2\odot \mathbf{G}$ & \hspace{2.5cm} $R_3\odot \mathbf{G}$ &  \hspace{2.5cm}$R_4\odot \mathbf{G}$  \\ \hline
 $\begin{bmatrix}
      1  &   1   &  1 &    1   &  1 &    1   &  1 &    1\\
     1  &  -1   & -1   &  1    & 1   & -1   & -1   &  1\\
     1   & -1    & 1   & -1    & 1   & -1   &  1   & -1\\
    -1    &-1    & 1   &  1    &-1   & -1   &  1   &  1\\
     1     &1   &  1   &  1    &-1   & -1   & -1   & -1\\
     1   & -1   & -1    & 1   & -1   &  1   &  1   & -1\\
     1   & -1  &   1    &-1  &  -1    & 1  &  -1   &  1\\
    -1   & -1 &    1    & 1 &    1    & 1 &   -1    &-1\\

 \end{bmatrix}$ &  $\begin{bmatrix}
  1 & -1 & -1 & 1 & 1 & -1 & -1 & 1 \\
1 & 1 & 1 & 1 & 1 & 1 & 1 & 1 \\
1 & 1 & -1 & -1 & 1 & 1 & -1 & -1 \\
-1 & 1 & -1 & 1 & -1 & 1 & -1 & 1 \\
1 & -1 & -1 & 1 & -1 & 1 & 1 & -1 \\
1 & 1 & 1 & 1 & -1 & -1 & -1 & -1 \\
1 & 1 & -1 & -1 & -1 & -1 & 1 & 1 \\
-1 & 1 & -1 & 1 & 1 & -1 & 1 & -1 \\

 \end{bmatrix}$ & $\begin{bmatrix}
     1 & -1 & 1 & -1 & 1 & -1 & 1 & -1 \\
1 & 1 & -1 & -1 & 1 & 1 & -1 & -1 \\
1 & 1 & 1 & 1 & 1 & 1 & 1 & 1 \\
-1 & 1 & 1 & -1 & -1 & 1 & 1 & -1 \\
1 & -1 & 1 & -1 & -1 & 1 & -1 & 1 \\
1 & 1 & -1 & -1 & -1 & -1 & 1 & 1 \\
1 & 1 & 1 & 1 & -1 & -1 & -1 & -1 \\
-1 & 1 & 1 & -1 & 1 & -1 & -1 & 1 \\

 \end{bmatrix}$ & $\begin{bmatrix}
     -1 & -1 & 1 & 1 & -1 & -1 & 1 & 1 \\
-1 & 1 & -1 & 1 & -1 & 1 & -1 & 1 \\
-1 & 1 & 1 & -1 & -1 & 1 & 1 & -1 \\
1 & 1 & 1 & 1 & 1 & 1 & 1 & 1 \\
-1 & -1 & 1 & 1 & 1 & 1 & -1 & -1 \\
-1 & 1 & -1 & 1 & 1 & -1 & 1 & -1 \\
-1 & 1 & 1 & -1 & 1 & -1 & -1 & 1 \\
1 & 1 & 1 & 1 & -1 & -1 & -1 & -1 \\

 \end{bmatrix}$ \\ \hline

 \hspace{2.5cm}$R_5\odot \mathbf{G}$ &  \hspace{2.5cm}$R_6\odot \mathbf{G}$ & \hspace{2.5cm} $R_7\odot \mathbf{G}$ &  \hspace{2.5cm}$R_8\odot \mathbf{G}$ \\ \hline
$\begin{bmatrix}
    1 & 1 & 1 & 1 & -1 & -1 & -1 & -1 \\
1 & -1 & -1 & 1 & -1 & 1 & 1 & -1 \\
1 & -1 & 1 & -1 & -1 & 1 & -1 & 1 \\
-1 & -1 & 1 & 1 & 1 & 1 & -1 & -1 \\
1 & 1 & 1 & 1 & 1 & 1 & 1 & 1 \\
1 & -1 & -1 & 1 & 1 & -1 & -1 & 1 \\
1 & -1 & 1 & -1 & 1 & -1 & 1 & -1 \\
-1 & -1 & 1 & 1 & -1 & -1 & 1 & 1 \\

\end{bmatrix}$ & $\begin{bmatrix}
    1 & -1 & -1 & 1 & -1 & 1 & 1 & -1 \\
1 & 1 & 1 & 1 & -1 & -1 & -1 & -1 \\
1 & 1 & -1 & -1 & -1 & -1 & 1 & 1 \\
-1 & 1 & -1 & 1 & 1 & -1 & 1 & -1 \\
1 & -1 & -1 & 1 & 1 & -1 & -1 & 1 \\
1 & 1 & 1 & 1 & 1 & 1 & 1 & 1 \\
1 & 1 & -1 & -1 & 1 & 1 & -1 & -1 \\
-1 & 1 & -1 & 1 & -1 & 1 & -1 & 1 \\

\end{bmatrix}$  & $\begin{bmatrix}
    1 & -1 & 1 & -1 & -1 & 1 & -1 & 1 \\
1 & 1 & -1 & -1 & -1 & -1 & 1 & 1 \\
1 & 1 & 1 & 1 & -1 & -1 & -1 & -1 \\
-1 & 1 & 1 & -1 & 1 & -1 & -1 & 1 \\
1 & -1 & 1 & -1 & 1 & -1 & 1 & -1 \\
1 & 1 & -1 & -1 & 1 & 1 & -1 & -1 \\
1 & 1 & 1 & 1 & 1 & 1 & 1 & 1 \\
-1 & 1 & 1 & -1 & -1 & 1 & 1 & -1 \\

\end{bmatrix}$ & $\begin{bmatrix}
    -1 & -1 & 1 & 1 & 1 & 1 & -1 & -1 \\
-1 & 1 & -1 & 1 & 1 & -1 & 1 & -1 \\
-1 & 1 & 1 & -1 & 1 & -1 & -1 & 1 \\
1 & 1 & 1 & 1 & -1 & -1 & -1 & -1 \\
-1 & -1 & 1 & 1 & -1 & -1 & 1 & 1 \\
-1 & 1 & -1 & 1 & -1 & 1 & -1 & 1 \\
-1 & 1 & 1 & -1 & -1 & 1 & 1 & -1 \\
1 & 1 & 1 & 1 & 1 & 1 & 1 & 1 \\

\end{bmatrix}$ \\ \hline

\end{tabular}}\label{table11}
\end{threeparttable}
\end{table*}

{
\subsection{Proposed GCS Construction}
In this sub-section, a construction of GCSs is established utilizing circulant matrices, GCPs, and GCPs' mates. }

\begin{theorem}\label{Theorem9}
Let $(\mathbf{a},\mathbf{b})$ be a GCP of length $N$  generated using {{Lemma \ref{lemmapre2}}} and let $(\mathbf{c},\mathbf{d})$ be its complementary mate. Let $\mathbf{A}=Cir(\mathbf{a})$, $\mathbf{B}=Cir(\mathbf{b})$, $\mathbf{C}=Cir(\mathbf{c})$, and $\mathbf{D}=Cir(\mathbf{d})$ be circulant matrices of size $N\times N$. The set of row sequences of the truncated matrix $\mathbf{G}^k$, defined as
\begin{equation}\label{matrixforccc1}
    \mathbf{G} = \begin{bmatrix}
        \mathbf{A} &  \mathbf{B}\\
         \mathbf{C} &  \mathbf{D}\\
    \end{bmatrix},
\end{equation}
forms a $(2N,2N-k)$-GCS, where $N=2^a10^b26^c$, $a,b,c \geq0$, and $0\leq k \leq 2N-2$.
\end{theorem}
\begin{IEEEproof}
    See Appendix F.
\end{IEEEproof}
\begin{remark}
    This approach is designed to achieve reduced computational complexity while maintaining a significantly larger flock size than previous constructions. The structural properties of this construction also distinguish it from our previous construction of GCSs, as it avoids the generation of CZCSs.
\end{remark}
\begin{corollary}\label{Corollary7}
    The matrix $\mathbf{G}=\begin{bmatrix}
         \mathbf{A} &  \mathbf{B}\\
         \mathbf{C} &  \mathbf{D}\\
    \end{bmatrix}$ forms a Hadamard matrix of order $2N$.
\end{corollary}
\begin{IEEEproof}
   See Appendix G.
\end{IEEEproof}

\begin{example}
    Let $\mathbf{a}=\left(1, 1, 1, -1\right)$ and $\mathbf{b}=\left(1, 1, -1, 1\right)$ be a GCP of length $4$, and let the pair $\mathbf{c}=\left(1,-1,1,1\right)$ and $\mathbf{d}=\left(1,-1,-1,-1 \right)$ be a complementary mate of $(\mathbf{a}, \mathbf{b})$. Then the rows of the matrix $\mathbf{G}^k$ form a $(8,8-k)$ -GCS for $0\leq k \leq 6$. For $k=0$, the matrix is
   \begin{equation*}
   \mathbf{G}=
\begin{bmatrix}   
1&	-1&	1	&1	&1	&1	&-1&	1\\
1&	1	&-1	&1&	1&	1&	1&	-1\\
1&	1&	1&	-1&	-1&	1&	1	&1\\
-1	&1&	1	&1	&1	&-1	&1	&1\\
1&	1&	1&	-1&	1&	-1&	-1&	-1\\
-1&	1&	1&	1&	-1&	1&	-1&	-1\\
1&	-1&	1&	1&	-1&	-1&	1&	-1\\
1&	1&	-1&	1&	-1	&-1	&-1	&1\\
\end{bmatrix}.
\end{equation*}
This matrix is also a Hadamard matrix of order $8$, i.e. $\mathbf{G} \mathbf{G}^\top=8I_8$.
\end{example}
\begin{remark}
   { It is important to distinguish the proposed framework from existing periodic constructions, such as the methodology established in \cite{Dok2015}. Although the approach in \cite{Dok2015} utilizes the transposes of two circulant matrices, $\mathbf{A}$ and $\mathbf{B}$, to satisfy the Hadamard condition, the current framework avoids the use of transposition. Instead, the proposed architecture is based on the algebraic relationship between an aperiodic GCP and its complementary mate. Using four distinct circulant blocks derived from this mate relationship, the framework establishes a unique construction path. This mate-based architecture provides enhanced structural flexibility, allowing the systematic generation of GCSs and Hadamard matrices of order $2N$ (for $N=2^a 10^b 26^c$) directly from aperiodic sequences.}
\end{remark}
\begin{remark}\label{Remark3}
   It should be noted from {{\textit{Theorem} \ref{Theorem9}}} that for $k=0$, we obtain a $(2N,2N)$-GCS, which is also a Hadamard matrix of order $2N$, where $N=2^a10^b26^c$ and $a,b,c\geq0$. The relationship is shown in Fig. \ref{fig:venn_intersection}.
\end{remark}
\begin{remark}
    {A key characteristic of our framework is the inherent trade-off between sequence length and flock size. The maximum sequence length is upper-bounded by the number of sequences in the set (e.g., a length of $2N-k$ requires a flock of size $2N$). This practical constraint means that the generation of very long sequences requires a large set, which should be considered when evaluating the framework for specific applications.}
\end{remark}

\begin{figure*}[ht!]
    \centering
    \begin{minipage}[b]{0.48\textwidth}
        \centering
\begin{tikzpicture}[
    set_czcss/.style={circle, draw=blue, fill=blue!20, fill opacity=0.5, minimum size=3cm, line width=1pt},
    set_ccc/.style={circle, draw=green!60!black, fill=green!20, fill opacity=0.5, minimum size=3cm, line width=1pt}
]
    \node[set_czcss, label={[blue, font=\bfseries]135:CZCSS}] (CZCSS) at (0,0) {};
    \node[set_ccc, label={[green!50!black, font=\bfseries]45:CCC}] (CCC) at (1.5,0) {};

    {\color{black}
    \node (label) at (4, 0.3) [font=\small\bfseries, text width=7em, text centered] { Optimal Construction (Thm. 5)};
    \draw[->, thick, shorten >=2pt] (label.south) -- (0.75, 0);
    }
\end{tikzpicture}
        \label{fig:venn_hierarchy}
    \end{minipage}
    \hfill 
    \begin{minipage}[b]{0.48\textwidth}
        \centering
        \begin{tikzpicture}
            \draw[blue, fill=blue!20, fill opacity=0.5] (0,0) circle (2cm);
            \draw[red, fill=red!20, fill opacity=0.5] (1.5,0) circle (2cm);
            \draw[green, fill=green!20, fill opacity=0.5] (0.75, -1.5) circle (1.8cm);
            \node at (0, 2.3) [blue, font=\bfseries] {CZCS};
            \node at (1.5, 2.3) [red, font=\bfseries] {GCS};
            \node at (0.75, -3.6) [green!50!black, font=\bfseries] {Hadamard};
            \node (label) at (0.75, -0.5) [font=\small\bfseries, text width=7em, text centered] { Proposed Constructions (Thm. 2)};
        \end{tikzpicture}
       \caption{ {Conceptual relationships between different code and matrix families. (Left) The intersection of CCCs and CZCSSs, highlighting that our proposed optimal construction lies in the overlapping region. (Right) The intersection of GCS, CZCSS, and Hadamard matrix properties, illustrating that our other constructions also reside in the region common to all three sets.}}
        \label{fig:venn_intersection}
    \end{minipage}
\end{figure*}

\subsection{Proposed Construction of CCC}
In this sub-section, we propose the construction of CCCs by extending the results for GCSs.

\begin{figure}
    \centering
    \includegraphics[width=0.8\linewidth]{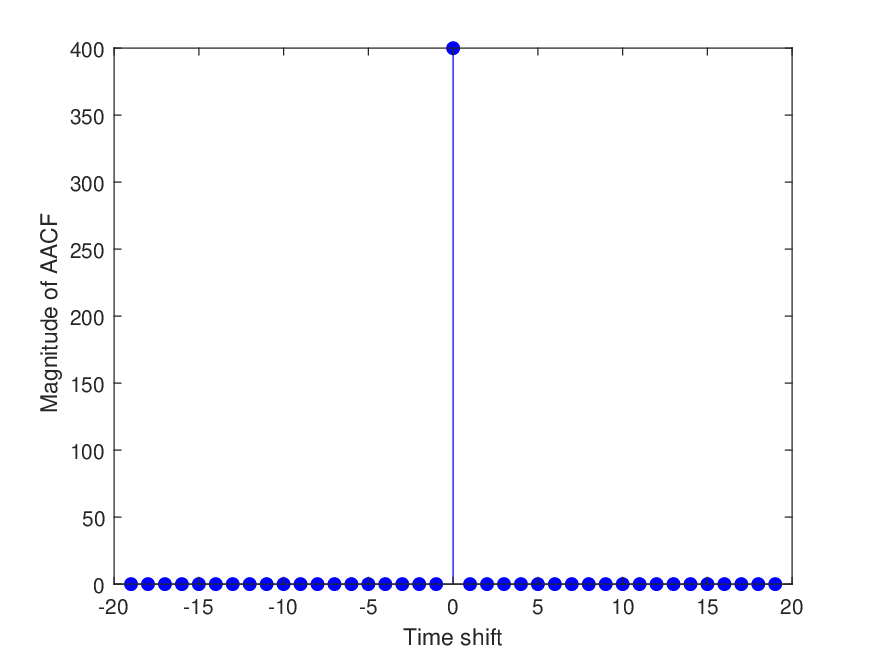}
    \caption{Sum of AACFs of each row of $R_i\odot\mathbf{G}$.}
    \label{fig1}
\end{figure}

\begin{figure}
    \centering
    \includegraphics[width=0.8\linewidth]{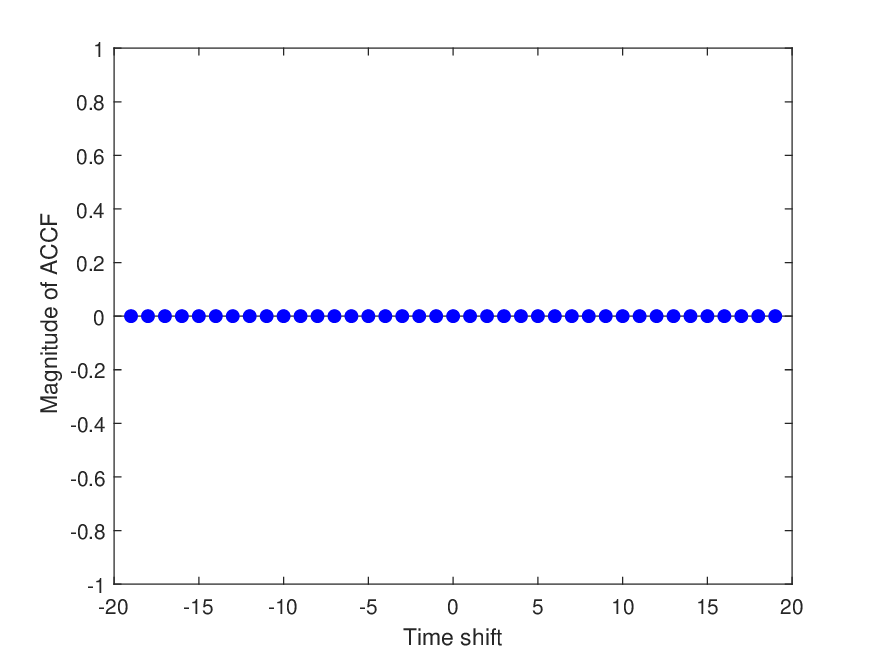}
    \caption{Sum of ACCFs between rows of $R_i\odot\mathbf{G}$ and $R_j\odot\mathbf{G}$.}
    \label{fig2}
\end{figure}

\begin{theorem}\label{Theorem10}
Let $R_{i}$ be the $i$-th row of the matrix $\mathbf{G}=\begin{bmatrix}
    \mathbf{A} & \mathbf{B}\\
    \mathbf{C} & \mathbf{D}\\
\end{bmatrix}$ from (\ref{matrixforccc1}), where $1\leq i \leq 2N$. Then the collection of sequence sets $\bigcup_{i=1}^{2N}R_i \odot \mathbf{G}$ forms a $(2N,2N,2N)$-CCC, where $\odot$ denotes the element-wise product.

\end{theorem}
\begin{IEEEproof}
   See Appendix H.
\end{IEEEproof}
{{\begin{corollary}\label{coro_ccc_hadamard}
If the matrix $\mathbf{G}$ in \textit{Theorem} \ref{Theorem10} is replaced by any real Hadamard matrix $\mathbf{H}$ of order $4n$, where $n \ge 1$ is an integer, then the collection of sequence sets $\bigcup_{i=1}^{4n}R_i \odot \mathbf{H}$ forms a binary $(4n,4n,4n)$-CCC, where $R_i$ is the $i$-th row of $\mathbf{H}$.
\end{corollary}}}
{\color{red}\begin{remark}\label{RemarkCCC}
    If the matrix $\mathbf{G}$ in \textit{Theorem} \ref{Theorem10} is replaced by any complex Hadamard matrix $\mathbf{H}$ of order $m$, where $n \ge 1$ is an integer, then the collection of sequence sets $\bigcup_{i=1}^{4n}R_i \odot \mathbf{H}$ forms a binary $(m,m,m)$-CCC, where $R_i$ is the $i$-th row of $\mathbf{H}$.
\end{remark}}
\begin{IEEEproof}
See Appendix I.
\end{IEEEproof}
\begin{remark}
    The above result can be extended to non-binary CCC of the parameter $(2m,2m,2m)$, where $2m$ is the order of complex Hadamard matrices and $m\geq1$.
\end{remark}
\begin{example}
Consider $\mathbf{a}=(1, 1, -1, 1, -1, 1, -1, -1, 1, 1)$ and $\mathbf{b}=(1, 1, -1, 1, 1, 1, 1, 1, -1, -1)$ to be a GCP of length 10, and let $\mathbf{c}=(-1,-1,1,1,1,1,1,-1,1,1)$ and $\mathbf{d}=(-1,-1,1,1,-1,1,-1,1,-1,-1)$ be its complementary mate. Construct the matrix $20 \times 20$.
\begin{equation*}
  \mathbf{G} = \begin{bmatrix}
        \mathbf{A} &  \mathbf{B}\\
         \mathbf{C} &  \mathbf{D}\\
    \end{bmatrix}.
\end{equation*}
Then $\bigcup_{i=1}^{20}R_i \odot \mathbf{G}$ forms a $(20,20,20)$-CCC, where $R_i$ denotes the $i$-th row of $\mathbf{G}$. The sum of AACFs for each set $R_i\odot \mathbf{G}$ is shown in Fig. \ref{fig1}. The sum of ACCFs between sets $R_i\odot \mathbf{G}$ and $R_j\odot \mathbf{G}$ is shown in Fig. \ref{fig2}.

\end{example}


\subsection{Proposed Construction of CZCSS}
In this sub-section, we propose the construction of optimal CZCSSs by extending the results for CZCSs.
\begin{theorem}\label{Theorem11}
    {Let $\mathbf{E}_{2^{n+2}}$ be the Hadamard matrix of order $2^{n+2}$ constructed recursively as defined in \textit{Theorem} \ref{Theorem8}.} Let $R_{i}$ be the $i$-th row of matrix $\mathbf{E}_{2^{n+2}}$, where $1\leq i \leq 2^{n+2}$. Then {the collection of code sets $\bigcup_{i=1}^{2^{n+2}}R_i \odot \mathbf{E}_{2^{n+2}}$} forms a $(2^{n+2},2^{n+2},2^{n+2},2^{n+1})$-CZCSS and also a $(2^{n+2},2^{n+2},2^{n+2})$-CCC for $n\geq1$.  
\end{theorem}
\begin{IEEEproof}
    See Appendix J.
\end{IEEEproof}

{
\begin{theorem}\label{Theorem12_GeneralizedCZCSS}
    Let $\mathbf{H}_{4n}$ be any Hadamard matrix of order $4n$, where $n \geq 1$. Define the recursive Hadamard matrix $\mathbf{H}_{8n} = \mathbf{H}_2 \otimes \mathbf{H}_{4n}$. Let $R_{i}$ denote the $i$-th row of $\mathbf{H}_{8n}$, where $1 \leq i \leq 8n$. Then, the collection of code sets is defined by
    \begin{equation}
        \mathcal{S} = \bigcup_{i=1}^{8n} R_i \odot \mathbf{H}_{8n}
    \end{equation}
    forms an optimal binary $(8n, 8n, 8n, 4n)$-CZCSS. Furthermore, this collection also constitutes an $(8n, 8n, 8n)$-CCC.
\end{theorem}
\begin{IEEEproof}
    See Appendix K.
\end{IEEEproof}

{\color{red}\begin{remark}\label{RemarkCZCSS}
    The construction in \textit{Theorem} \ref{Theorem12_GeneralizedCZCSS} can be extended to generate non-binary sequence sets. Given a complex Hadamard matrix $\mathbf{H}_{m}$ of order $m$, where $m \geq 1$, let $\mathbf{H}_{2m} = \mathbf{H}_2 \otimes \mathbf{H}_{m}$. The collection of sets formed by the element-wise product of the rows of $\mathbf{H}_{2m}$ results in a non-binary (complex) $(2m, 2m, 2m, m)$-CZCSS, where $m\geq1$. 
\end{remark}}
}
\begin{example}\label{Example8}
{Let us construct the order-8 Hadamard matrix $\mathbf{E}_8$ as defined in \textit{Theorem} \ref{Theorem8}. Let $\mathbf{E}_4$ be a circulant Hadamard matrix of order 4:}
\begin{equation*}
    \mathbf{E}_4=\begin{bmatrix}
        1 & -1 & 1 & 1\\
        1 & 1 &-1 & 1\\
        1 & 1 & 1 & -1\\
        -1 & 1 & 1 & 1\\
    \end{bmatrix}, 
\quad \text{and} \quad \mathbf{E}_8=\begin{bmatrix}
    \mathbf{E}_4 & \mathbf{E}_4\\
    \mathbf{E}_4 & -\mathbf{E}_4\\
\end{bmatrix}.
\end{equation*}
Then $\bigcup_{i=1}^{8}R_i \odot \mathbf{E}_8$ forms an $(8,8,8,4)$-CZCSS, where $R_i$ denotes the $i$-th row of $\mathbf{E}_8$. Each code set is presented in Table \ref{table11}.
\end{example}
\begin{remark}
    {A key feature of the proposed construction is its optimality \cite{EnhancedZLiu}. This means that the set is not only a $(2^{n+2}$, $2^{n+2}$, $2^{n+2}$, $2^{n+1})$-CZCSS but also concurrently a perfect binary $(2^{n+2}, 2^{n+2}, 2^{n+2})$-CCC for $n\geq1$. This dual nature is depicted in Fig.~\ref{fig:venn_intersection}.}
\end{remark}
\begin{remark}
 All proposed constructions also provide $q$-phase CZCSSs, GCSs, CCCs, and CZCSSs, where $q$ is a positive integer. 
\end{remark}

\begin{figure}[t]
\centering
\begin{tikzpicture}[
    block/.style={rectangle, draw, fill=blue!5, text width=7.5em, text centered, rounded corners, minimum height=3.5em, font=\small\bfseries},
    line/.style={draw, -stealth, thick},
    label_style/.style={font=\scriptsize, align=center}
]

\node [block] (seed) at (0,0) {Hadamard Seed \\ $\mathbf{H}$};
\node [block] (exp) at (5,0) {Expanded Matrix \\ $\mathbf{H}_{8n} = \mathbf{H}_2 \otimes \mathbf{H}$};
\node [block] (czcs) at (10,0) {CZCS / GCS};
\node [block] (ccc) at (0,-3.5) {Binary \\ CCC};
\node [block] (czcss) at (5,-3.5) {Optimal \\ CZCSS};

\draw [line] (seed) -- (exp);
\draw [line] (exp) -- (czcs);
\draw [line] (seed) -- (ccc);
\draw [line] (exp) -- (czcss);

\node [label_style] at (2.5, 0.5) {Kronecker \\ Expansion ($\otimes$)};
\node [label_style] at (7.5, 0.5) {Systematic \\ Truncation};
\node [label_style, rotate=-90] at (0.4, -1.5) {Row-wise ($\odot$)};
\node [label_style, rotate=-90] at (5.4, -1.5) {Row-wise ($\odot$)};

\end{tikzpicture}
\caption{{A systematic construction path showing the transformation of a single Hadamard matrix into various complementary structures.}}
\label{fig:systematic_diagram}
\end{figure}
\subsection{Classification of GCSs and Their Relation with Hadamard Matrices}
In this section, we classify GCSs based on their properties and show the relationship between GCSs and Hadamard matrices.
\begin{itemize}
    \item A square matrix $\mathbf{G}$ of order $n$ is called a type-1 GCS if it is a Hadamard matrix. That is, the rows of $\mathbf{G}$ form an $(n,n)$-GCS and $\mathbf{G}\mathbf{G}^\top=nI_n$, where $n$ is $1,2$ or a multiple of $4$.
\end{itemize}

\begin{example}
 Let us take
    \begin{equation}     
   \mathbf{G}=\begin{bmatrix}
        -1 & 1 & 1 & 1\\1 &-1&1 &1 \\1 & 1&-1 &1\\1 & 1&1 &-1\\
    \end{bmatrix}. \end{equation} 
    The rows of $\mathbf{G}$ form a $(4,4)$-GCS and $\mathbf{G}\mathbf{G}^\top=4I_4$, so $\mathbf{G}$ is a type-1 GCS.
\end{example}
\begin{itemize}
\item A square matrix $\mathbf{G}$ of order $n$ is called a type-2 GCS if its rows form a $(n,n)$ -GCS but it is not a Hadamard matrix, i.e., $\mathbf{G}\mathbf{G}^\top \neq nI_n$.
\end{itemize}
\begin{theorem}\label{Theorem12}
    The set of rows of any Hadamard matrix of order $n$ forms an $(n,n)$-GCS, where $n$ is $1$, $2$, or a multiple of $4$.
\end{theorem}
\begin{IEEEproof}
    See Appendix L.
\end{IEEEproof}
\begin{remark}
    The converse of {{\textit{Theorem} \ref{Theorem12}}} is not true.
    \begin{example}
       Consider the $(8,8)$-GCS generated from \cite{Wang2020SPL}:
       \begin{equation}
           A=\begin{bmatrix}
              1 &1 &1 &-1& -1& 1& 1& 1\\
    1& 1 &-1& 1& 1& -1& 1& 1\\
    -1& -1 &-1 &1& 1 &1 &-1& 1\\
    -1& -1& 1& -1& 1& 1& 1& -1\\
    -1& -1& -1& 1& -1& 1& 1& 1\\
    -1 &-1& 1& -1& 1& -1& 1& 1\\
    1 &1& 1& -1& 1& 1& -1& 1\\
    1 &1 &-1& 1& 1& 1& 1& -1 \\
           \end{bmatrix}.
       \end{equation}
       The rows of the matrix above form an $(8,8)$-GCS, but it is not a Hadamard matrix of order $8$. This is an example of a type-2 GCS.
    \end{example}
\end{remark}
\begin{theorem}\label{Theorem13}
    From the collection of all matrices whose rows form an $(n,n)$-GCS, at least one is a Hadamard matrix of order $n$, where $n=1, 2$ or a multiple of $4$.
\end{theorem}
\begin{IEEEproof}
    See Appendix M.
\end{IEEEproof}

{
\subsection{Unified Transformation Pathway}
To provide a clearer understanding of the proposed work, we present a unified path for all constructions. As illustrated in Fig. \ref{fig:systematic_diagram}, our framework treats any standard Hadamard matrix as a flexible tool rather than a fixed sequence. By starting with a single seed, we can reach different types of sets—such as GCS, CZCS, CCC, and optimal CZCSS—using simple algebraic rules such as expansion and truncation. This path proves that all these complementary structures share a common origin in the orthogonality of Hadamard matrices.}

\begin{table*}[ht!]
\begin{threeparttable}[!]
\centering
\caption{Comparison of CZCS Constructions}
\resizebox{\textwidth}{!}{
\begin{tabular}{|l|l|l|l|l|l|}
\hline 
         Ref.    &Method      &Set size  &Length &Zone        & Constraint  \\ \hline  
        \cite{liu2020cross} &Indirect &$M$ &$L$ &$Z$ &Requires existing $(L,Z)$-CZCP \\ \hline
        \multirow{5}{*}{\cite{Huang2022}} &  \multirow{5}{*}{Indirect}  &  \multirow{5}{*}{4} &$N$ &$N$ &Requires existing GCP of length $N$ \\
        \cline{4-6} & & &$N_1+N_2$ &$min(N_2,N_1+Z_2)$ & Requires existing GCP of length $N_1$ and $(N_2,Z_2)$-CZCP\\
        \cline{4-6} & & &$2L$ &$L$ & Requires existing $(4,L)$-GCS\\
        \cline{4-6} & & &$2N$ &$2N$ & Requires existing GCP of length $N$ \\
        \cline{4-6} & & &$4L$ &$2L$ & Requires existing $(4,L)$-GCS\\ \hline
        \cite{ChenISITCZCS} &Indirect &$4$ &$N$ &$N$ &Requires existing GCP of length $N$ \\ \hline
        \multirow{3}{*}{\cite{shibsankarCZCS}} &\multirow{3}{*}{Indirect}  &\multirow{3}{*}{4} &$3L$ &$L$ &\multirow{3}{*}{$L$ is the length of a seed GCP}\\
        \cline{4-5} & & &$7L$ &$2L$ & \\
         \cline{4-5} & & &$14L$ &$6L$ & \\ \hline
         \cite{PraveenCZCS} &GBF &$2^{k+1}$ &$2^{m-1}+2^{\delta}$ &$2^{pi_k(1)-1}+2^{\delta}$ &$m\geq4, 0\leq \delta \leq m-1$ \\ \hline
         \multirow{2}{*}{\cite{HuangCZCS}} &\multirow{2}{*}{GBF} &$2^{k+1}$ &$2^{m-k}(2^k-1)+2^v$ &$2^{m-1}$ &$m\geq2,2\leq k \leq m$ \\
         \cline{3-6} & &$2^{k_1+2}$ &$L_{H}$ &$Z_{H}$ &$m\geq2,1\leq k_1 \leq m$ \\ \hline
        \textit{Theorem} \ref{Theorem8} &Circulant Hadamard matrices &$2^{n+2}$ &$2^{n+2}-k$ &$2^{n+1}-(k-2^n)\lfloor \frac{k}{2^n} \rfloor$ &$n\geq1, 0\leq k \leq 2^{n+1}-1$ \\ \hline
        {\textit{Theorem} \ref{Theorem9_GeneralizedCZCS}} & { Hadamard matrices} & {$8n$} & {$8n-k$} & {$4n-k$} & {$n \geq 1, 0 \leq k \leq 4n-1$} \\ \hline
        {\color{red}\textit{Remark} \ref{RemarkCZCS}} & {Complex Hadamard matrices} & {$2m$} & {$2m-k$} & {$m-k$} & {$m \geq 1, 0 \leq k \leq m-1$} \\ \hline
\end{tabular}}\label{table2}
\begin{tablenotes}
      \small
      \item For \cite{HuangCZCS}, $L_{H} = Z_{H} = 2^{m-1}+\sum_{\beta=1}^{k_1-1} a_{\beta} 2^{\pi(m-k_1+\beta)-1} + 2^{v_1}$, where $a_\beta \in \{0, 1\}$.
    \end{tablenotes}
\end{threeparttable}
\end{table*}
\begin{table*}[ht!]
\begin{threeparttable}[t]
\centering
\caption{Comparison of Hadamard Matrix Constructions}
\resizebox{\textwidth}{!}{
\begin{tabular}{|l|l|l|l|l|}
\hline 
         Ref.    &Method      &Order         & Constraint  \\ \hline  
        \cite{sylvester1867thoughts} & Adjoining matrices & $2n$ & $n$ is the order of a Hadamard matrix \\ \hline 
        \cite{paley1933orthogonal} &Quadratic residue in finite field & $p+1$ and $2(p+1)$ & $p$ is a prime where $p\equiv 3 \pmod{4}$ or $p\equiv 1 \pmod{4}$ \\ \hline
        \cite{BaumertHall1965} & Williamson-type matrices &$12t$ &$t$ is a positive integer \\ \hline 
        \cite{CooperWallis1972} &Recursive approach &$4t$ & $t\in\{1,3,5,7,...,19\}$ \\ \hline
        \cite{turyn1970complex} &Complex Hadamard matrices &$8nh$ & $4h$ and $2n$ are orders of real and complex Hadamard matrices, respectively \\ \hline
        \cite{Wallis1973} &T-matrices & $28m$, $36m$, and $44m$ & $m$ is the order of T-matrices \\ \hline
        \cite{KoukouvinosKounias1989} & T-matrices &$2^tq$ &$q$ is the sum of the lengths of two GCPs and {{$t\in \mathbb{Z}^+$}} \\ \hline
        \textit{Remark} \ref{Remark2} & Circulant matrices &$2N$ &$N=2^a10^b26^c$, where $a,b, c \geq 0$ \\ \hline
\end{tabular}}\label{table1}
\end{threeparttable}
\end{table*}
\section{Comparison}
This section compares the proposed constructions with the current state-of-the-art.
\begin{itemize}
  \item  A complete comparison of Hadamard matrix constructions is provided in Table \ref{table1}. {Historically, the construction of Hadamard matrices has relied on specific algebraic structures such as quadratic residues \cite{paley1933orthogonal}, Williamson-type matrices \cite{sylvester1867thoughts,BaumertHall1965,goethals1967orthogonal}, and T-matrices \cite{Wallis1973,CooperWallis1972,turyn1970complex}. A common challenge with many of these classical methods is the requirement for specific seed matrices (e.g., Williamson-type or T-matrices), which are not known for all orders and can be difficult to find. In contrast, this paper presents two new systematic construction frameworks that leverage more readily available sequences. The proposed method, using circulant matrices seeded by GCPs, directly constructs matrices of order $2N$, where $N=2^{a}10^b26^c$. A key advantage of our GCP-based approach is its low complexity and high flexibility, as the required seed sequences can be easily generated for all binary GCP lengths, overcoming the availability issues of earlier methods.}
  
{
\item  A comprehensive comparison of CZCS constructions is summarized in Table \ref{table2}. While the GBF-based approach in \cite{HuangCZCS} achieves arbitrary length coverage with a power-of-two flock size, the framework established in this work provides a fundamentally different algebraic methodology. By utilizing circulant Hadamard matrices as foundational seeds, Theorem \ref{Theorem8} provides a systematic path to construct CZCSs of arbitrary lengths. Furthermore, the generalized framework in Theorem \ref{Theorem9_GeneralizedCZCS} employs arbitrary binary or complex Hadamard matrices to generate CZCSs with flexible, non-power-of-two flock sizes. Critically, this matrix-based approach breaks the flock-size constraints ($M=2^k$) inherent in existing Boolean-based designs \cite{HuangCZCS}, allowing for user capacities that are not restricted to powers of two. }


\begin{table*}[ht!]
\begin{threeparttable}[t]
\centering
\caption{Comparison of GCS Constructions}
\resizebox{\textwidth}{!}{
\begin{tabular}{|l|l|l|l|l|l|}
\hline 
         Ref.    &Method      &Length  &Set Size   &Phase     & Constraint  \\ \hline  
         \cite{paterson2000generalized} &GBF &$2^m$ &$2^m$ &$q$ &$m\geq1$,  $2 | q$  \\ \hline
          \cite{schmidt2007complementary} &GBF &$2^m$ &$2^m$ & $q$&$m\geq1$,  $2 | q$  \\ \hline
         \multirow{2}{*}{\cite{chen2016complementary}} &\multirow{2}{*}{GBF} & $2^{m-1}+2^v$ &$4$ &\multirow{2}{*}{$q$} &\multirow{2}{*}{$1\leq v \leq m-1$, $m\geq2$,  $2 | q$ }\\
         \cline{3-4} & & $2^{m-1}+1$ & $2^{k+1}$ & & \\ \hline
         \cite{wang2017method} &PU matrices &$2^{m-1}+2^v$ &$2^{k+1}$ &$q$ &$1\leq v \leq m-1$, $m\geq2$,  $2 | q$  \\ \hline
         \cite{chen2017new} &GBF &$2^{m-1}+2^v$ &$2^{k+1}$ &$q$ &$1\leq v \leq m-1$, $m\geq2$,  $2 | q$  \\ \hline
         \cite{chen2018novel} &GBF &$2^{m-1}+\sum_{\alpha=1}^{k-1}{a_{\alpha}2^{\pi(m-k+\alpha)-1}}+2^{v}$& $2^{k+1}$  &$2$ & $k,m\geq2$, $v\geq0$ \\ \hline
         \multirow{2}{*}{\cite{adhikary2019new}} &\multirow{2}{*}{GBF}  &$N+1$, $N+2$ &$4$ &\multirow{2}{*}{$q$} &\multirow{2}{*}{$N=2^{a}10^b26^c$, $a,b,c\geq0$,  $2 | q$ } \\
          \cline{3-4} & & $2N+3$ &$8$ & & \\ \hline 
          \multirow{2}{*}{\cite{Wang2020SPL}} &\multirow{2}{*}{Concatenation} &$N+M$ &$4$ &\multirow{2}{*}{$q$} &{$N=2^{a}10^b26^c$, $a,b,c\geq0$,  $2 | q$ } \\
          \cline{3-4} & & $N+P$ &$8$ & &$P$ is the length of a GCS of set size $4$  \\ \hline 
          {{\multirow{2}{*}{\cite{wang2021new}}}} & {{Paraunitary (PU) matrix}}  & $N^n$ & $N^m$  & {{$q$}} & {{$n,m>0$} and $N$ denotes the order of PU matrix} \\ \cline{2-6}
         & {{Paraunitary (PU) matrix}} & $2^m$ & $2^n$ & {{$q$}}  & {{$k,n,m>0$}} \\ \hline
          \cite{sheneven} &Indirect &$LN$ &$4$ &$q$ & $L$ is the length of an ESCP,  $2 | q$  \\ \hline
          {\cite{wang2023constructions}} & {Permutation polynomial }& {$2^m$}  & {$2^m$} & {$q$}  &{$m\geq1$ and $q$ is an even integer} \\ \hline
          \cite{chen2021GCS} &GBF &$2^{m-1}+2^t$ &$2^{k+1}$ &$q$ & $k\leq m-1$, $m\geq2$,  $2 | q$  \\ \hline
          \cite{abhishekGCS} &EBF &$L$ &$p^k$ & $q$ & $p,L \in \mathbb{N}$ and $p | q$ \\ \hline 
          \textit{Theorem} \ref{Theorem9} &Circulant matrix &$2N-k$ &$2N$ &$q$ & $N=2^{\alpha}10^\beta26^\gamma$,  $2| q$, $0\leq k \leq 2N-2$  \\ \hline
          \textit{Corollary} \ref{Corollary6} &Circulant Hadamard matrix &$2^{n+2}-k$ &$2^{n+2}$ &$q$ &$n\geq1$ and $0 \leq k \leq 2^{n+1}-2$,  $2 | q$  \\ \hline
\end{tabular}}\label{table3}
\end{threeparttable}
\end{table*}
\item A complete comparison of the GCS constructions is provided in Table \ref{table3}. Constructions of GCSs with lengths that are powers of two were presented in \cite{paterson2000generalized,schmidt2007complementary}, while non-power-of two lengths were addressed in \cite{chen2016complementary}. GBF-based constructions included GCSs of set size $4$ and length $2^{m-1}+2^v$ \cite{wang2017method}, and set size $2^{k+1}$ with the same length \cite{chen2017new}. More generalized lengths of the form $2^{m-1}+\sum_{\alpha=1}^{k-1}{a_{\alpha}2^{\pi(m-k+\alpha)-1}}+2^v$ with set size $2^{k+1}$ were introduced for $k,m\geq2$ and $v\geq0$ \cite{chen2018novel}. GCSs of lengths $N+1$, $N+2$, and $2N+3$ with set sizes $4$ and $8$ were constructed by extending GCPs \cite{Adhikary2019}. Further flexible constructions based on GCPs were presented in \cite{Wang2020SPL, Pai2020EL}. A construction using ESCPs yielded lengths $LN$ with a set size $4$ \cite{sheneven}. GBF-based GCSs of length $2^{m-1}+2^t$ and set size $2^{k+1}$ were developed for $k \leq m-1$ \cite{chen2021GCS}.{{ Algebraic advancements have established para-unitary (PU) matrices \cite{wang2017method, wang2021new} and permutation polynomials \cite{wang2023constructions} as powerful tools for generating GCSs with diverse parameters. While a recent construction based on GBF achieved coverage for all sequence lengths, it remains restricted by power of two flock size limitations \cite{abhishekGCS}. In this work, two novel GCS construction frameworks are established: the first utilizes circulant Hadamard matrices as recursive seeds, while the second is derived from circulant matrices seeded by GCPs and their aperiodic mates. Although alternative methods exist for constructing GCSs with flexible lengths, the proposed frameworks offer a distinct structural approach rooted in circulant operators. This systematic methodology reveals a direct algebraic link between GCS properties and Hadamard matrices, contributing to the diversity of available sequence design techniques. By utilizing aperiodic GCPs as foundational seeds, the proposed framework inherits the flexible parameters of these sequences. Instead of being restricted to a fixed design, this approach allows for the straightforward adjustment of flock size and phase simply by selecting an appropriate GCP seed from the wide variety of available lengths. It is important to note that these constructions involve an inherent trade-off between sequence length and flock size, as detailed in Remark 5.}}

{
\item A comparative analysis of CCC constructions is summarized in Table~\ref{table5}. The literature reveals a technical trade-off: inherently binary methods \cite{conCCCHM, Rathinakumar2008} are generally restricted to power-of-two flock sizes, while general PU-based frameworks \cite{concccPUmatrices, shibsankarvarious} involve high computational complexity. Crucially, the proposed methodology is fundamentally different from PU-based approaches. While PU theory relies on normalized Hadamard matrices to satisfy unitary conditions, our framework utilizes Hadamard matrices directly. By bypassing the normalization requirement, this approach ensures that the sequences remain strictly binary ($\pm 1$) while simultaneously breaking the power-of-two flock size limitations prevalent in existing works. This enables the construction of binary $(2N, 2N, 2N)$ and $(4n, 4n, 4n)$ CCCs for $n \ge 1$ using simple algebraic operations, where $N=2^a10^b26^c$ and $a,b,c\geq0$. The construction is also extended for complex or unimodular CCC of the form $(m,m,m)$, where $m\geq1$. A significant technical advantage of this method is that the resulting CCCs themselves are Hadamard matrices, ensuring excellent orthogonality and establishing a direct link between classical matrix theory and flexible code design.}

 \begin{table*}[ht!]
\begin{threeparttable}[t]
\centering
\caption{Comparison of Proposed and Previous CCC Constructions}
\resizebox{\textwidth}{!}{
\begin{tabular}{|l|l|l|l|l|}
\hline 
         Ref.    &Method/Tool &Phase &Parameter       & Constraint  \\ \hline
         \cite{concccunitarymatrices} &Unitary matrices &$q$ &$(M,M,M^N)$ & $q,N\geq2$ \\ \hline
         \cite{conCCCHM} &Hadamard matrices &$2$ &$(2^{N-r},2^{N-r},2^N)$ & $r=1,2,\cdots,N-1$ \\ \hline
         \cite{concccBooleanfunction} &Boolean function &$q$ &$(2^{k+1},2^{k+1},2^m)$ &$q\geq2$, $m,k \geq1$ and $k=m-1$ \\ \hline
         {{\cite{Rathinakumar2008}}} & {{Reed-Muller Codes}} &{{$2$}} &{{$(2^k, 2^k, 2^m)$}} & {{$m \geq 1, 0 \leq k \leq m$}} \\ \hline
          \cite{sarkar2021multivariable} & GBF & $\prod_{i=1}^{k}p^{n_i+1}_i$ &$\prod_{i=1}^{k}p^{m_i}_i$ &$q=\prod_{i=1}^{k}p_i$, $p_i,m_i\geq2$, $n_i\geq0$ and $p_i$ is a prime number\\ \hline
         
         \cite{concccPUmatrices} & PU matrices &$q$ &$(M,M,M^N)$ & $N \geq 0, q \geq 2$ \\ \hline
    
         {{\multirow{2}{*}{\cite{wang2021new}}}} & {{Paraunitary (PU) matrix}} & {{$q$}} & {{$(N, N^n, N^m)$}} & {{$n,m>0$} and $N$ denotes the order of PU matrix} \\ \cline{2-5}
         & {{Paraunitary (PU) matrix}} & {{$q$}} & {{$(2^k, 2^{kn}, 2^{km})$}} & {{$k,n,m>0$}} \\ \hline
      \cite{shibsankarvarious} &PU matrices &Q &$(M,M,M^jM'\prod_{j=0}^{J}{l_j}^{N_j}l^N)$ &$1\leq M'\leq M, l_j | M, l |M'$, $lcm\{q,q_p\}_{p=0}^{P-1}=Q$ \\ \hline
         \cite{wang2023constructions} & Permutation polynomial &$q$ &$(2^m,2^m,2^m)$ &$m\geq1$ and $q$ is an even integer \\ \hline
        
         \textit{Theorem} \ref{Theorem10} & Circulant matrix &$q$ &$(2N,2N,2N)$ &$N=2^a10^b26^c, a,b,c \geq0$, $2|q$ \\ \hline
          \textit{Corollary} \ref{coro_ccc_hadamard} &  Hadamard matrix &  $2$ &  $(4n,4n,4n)$ &  $n\geq1$ \\ \hline
          {\color{red}\textit{Remark} \ref{RemarkCCC}} & Complex Hadamard matrix & $q$ & $(m,m,m)$ & $m,q\geq1$ \\ \hline

\end{tabular}}\label{table5}
\end{threeparttable}
\end{table*}
\begin{table*}[ht!]
\begin{threeparttable}[t]
\centering
\caption{Comparison of Existing CZCSS Constructions with the Proposed Method}
\resizebox{\textwidth}{!}{
\begin{tabular}{|l|l|l|l|l|l|l|l|}
\hline 
         Ref.    &Method  &Code size    &Set size  &Length &Zone &Optimality       & Constraint  \\ \hline
         \cite{praveenCZCCS} &GBF &$2^{n+1}$ & $2^{n+1}$ &$2^{m-1}+2$ &$2^{\pi(m-3)}$ &No & $n\in \mathbb{Z}^+, m\geq 4$  \\ \hline
        \multirow{3}{*}{\cite{EnhancedZLiu}} & \multirow{2}{*}{Indirect}  &$M$ &$N$ &$2L$  &$Z$ &No &$(M,N,L,Z+1)$-ZCCS \\ \cline{3-7} & & $M$ &$N$ &$2L$ &$L$ &No &$(M,N,L)-MOGCS$ \\ \cline{2-7} 
        &GBF &$2^k$ &$2^v$ &$2^m$ & $2^{\pi_1(1)-1}$ &Conditional & $k,m,v \in\mathbb{Z}^+, v \leq k$ \\ \hline
         \textit{Theorem} \ref{Theorem11} &Circulant Hadamard matrices &$2^{n+2}$ &$2^{n+2}$ &$2^{n+2}$  &$2^{n+1}$ & Yes & $n\geq1$\\ \hline  
         {\textit{Theorem} \ref{Theorem12_GeneralizedCZCSS}} & { Hadamard matrices} & {$8n$} & {$8n$} & {$8n$} & {$4n$} & {Yes} & {$n \geq 1$} \\ \hline
         {\color{red}\textit{Remark} \ref{RemarkCZCSS}} & {Complex Hadadmard matrices} & {$2m$} & {$2m$} & {$2m$} & {$m$} & {Yes} & {$m \geq 1$} \\ \hline
\end{tabular}}\label{table4}
\end{threeparttable}
\end{table*}
\item  {A complete comparison of CZCSSs is provided in Table~\ref{table4}. The key limitation of existing methods based on GBFs \cite{praveenCZCCS, EnhancedZLiu} and concatenation \cite{EnhancedZLiu} is their inability to guaranty optimality in general parameters. These constructions are either explicitly non-optimal or achieve optimality only under highly restrictive conditions for only power-of-two length, resulting in suboptimal ZCZ. Our proposed framework overcomes these critical limitations by establishing a universal algebraic link between Hadamard matrices and optimal sequence sets. As demonstrated in Theorems \ref{Theorem11} and \ref{Theorem12_GeneralizedCZCSS}, we generate binary $(2^{n+2}, 2^{n+2}, 2^{n+2}, 2^{n+1})$-CZCSS and $(8n, 8n, 8n, 4n)$-CZCSS that achieve the maximum theoretical  $Z = NL/(2M)$, where $n\geq1$. Furthermore, our Remark 4 extends these results to the complex domain, providing $(2m, 2m, 2m, m)$-CZCSS from arbitrary complex Hadamard seeds, where $m\geq1$. This approach not only provides optimal parameters, but also reveals a valuable structural relationship, as these code sets concurrently satisfy the requirements for CCC. To our knowledge, this is the first generalized framework capable of converting any binary or complex Hadamard matrix into an optimal CZCSS.}

\end{itemize}

\section{Conclusion}

{
In this paper, we proposed a comprehensive framework rooted in the algebraic properties of circulant structures to construct multiple classes of complementary sets and Hadamard matrices. Circulant Hadamard matrices of order $4$ initiated a recursive construction that produced binary CZCSs and GCSs for power-of-two-based lengths and achieved a maximum ZCZ ratio of $2/3$ for CZCZ. Significantly, the recursive framework generalized to establish that any binary or complex Hadamard matrix enabled the construction of CZCSs of arbitrary length with non power-of-two flock size. Furthermore, a second distinct framework was proposed that uses circulant matrices seeded by GCPs to construct GCSs with flexible flock sizes without power of-two of the form $2N$, where $N=2^a10^b26^c$ for $a,b,c \ge 0$. A key finding confirmed that the GCSs generated from both methods functioned as Hadamard matrices. Finally, the foundational GCS and CZCS constructions extended to generate two important related families: flexible $(2N,2N,2N)$, $(4n,4n,4n)$, and $(m,m,m)$-CCCs, as well as optimal binary $(8n,8n,8n,4n)$-CZCSSs and their complex counterparts of the parameters $(2m,2m,2m,m)$, where $n,m \ge 1$ and $N=2^a10^b26^c$ for $a,b,c \ge 0$. A systematic transformation pathway established that these diverse complementary structures shared a common origin in Hadamard seeds. This unified framework, supported by a systematic diagram, provided a clear algebraic roadmap that simplified the design process and broke the traditional flock-size constraints inherent in combinatorial methods. Collectively, these contributions established a powerful and systematic link between circulant operators, various classes of complementary sets, and the theory of Hadamard matrices, and opened new avenues for optimal sequence design. Future research focused on exploring the complete theoretical relations among CZCSs, GCSs, and Hadamard matrices.
}

\begin{appendices} 

{\section{Proof of Lemma 4}
Let $\mathbf{a} = (a_0, a_1, \dots, a_{L-1})$ and $\mathbf{b} = (b_0, b_1, \dots, b_{L-1})$ be two sequences of length $L$. Based on the definition of the operator of the right cyclic shift $T_1$ in Section II, the columns $i$ -th and $j$ -th of the circulant matrices $\text{Cir}(\mathbf{a})$ and $\text{Cir}(\mathbf{b})$ correspond to the vectors $(T_1^i(\mathbf{a}))^\top$ and $(T_1^j(\mathbf{b}))^\top$, respectively. The elements of these vectors are:
\begin{align}
    T_1^i(\mathbf{a}) &= (a_{L-i}, a_{L-i+1}, \dots, a_{L-1}, a_0, a_1, \dots, a_{L-1-i}), \\
    T_1^j(\mathbf{b}) &= (b_{L-j}, b_{L-j+1}, \dots, b_{L-1}, b_0, b_1, \dots, b_{L-1-j}).
\end{align}
The inner product $\langle (T_1^{i}(\mathbf{a}))^\top, (T_1^{j}(\mathbf{b}))^\top \rangle$ is expressed as the matrix product $T_1^{i}(\mathbf{a}) (T_1^{j}(\mathbf{b}))^\dagger$, where $\dagger$ denotes the conjugate transpose. For $j \leq i$, let $k' = i - j$ denote the relative shift between the columns. The expansion of this product is as follows.
\begin{equation}\label{proof_expansion_sum}
    \begin{aligned}
       T_1^{i}(\mathbf{a}) (T_1^{j}(\mathbf{b}))^\dagger ={}& a_{L-i}b^*_{L-j} + a_{L-i+1}b^*_{L-j+1} + \dots + a_{L-i+j-1}b^*_{L-1} \\
       & + a_{L-i+j}b^*_0 + \dots + a_{L-1}b^*_{i-j-1} \\
       & + a_0b^*_{i-j} + a_1b^*_{i-j+1} + \dots + a_{L-1-i}b^*_{L-1-j}.
    \end{aligned}
\end{equation}
Rearranging the terms in (\ref{proof_expansion_sum}) to group the indices relative to the shift $k'$, we obtain:
\begin{equation}
    \begin{aligned}
    T_1^{i}(\mathbf{a}) (T_1^{j}(\mathbf{b}))^\dagger ={}& \underbrace{a_{L-k'}b_0^* + a_{L-k'+1}b_1^* + \dots + a_{L-1}b_{k'-1}^*}_{\text{Term 1}} \\
    & + \underbrace{a_0b_{k'}^* + a_1b_{k'+1}^* + \dots + a_{L-1-k'}b_{L-1}^*}_{\text{Term 2}}.
    \end{aligned}
\end{equation}
Applying the definition of ACCF from Eq. (\ref{eqn1}):
\begin{itemize}
    \item \textbf{Term 2} corresponds to the summation $\sum_{m=0}^{L-1-k'} a_m b^*_{m+k'}$, which is exactly $\mathcal{C}(\mathbf{a}, \mathbf{b})(k')$.
    \item \textbf{Term 1} corresponds to the summation $\sum_{m=0}^{k'-1} a_{L-k'+m} b^*_m$, which is exactly $\mathcal{C}(\mathbf{b}, \mathbf{a})(L - k')$.
\end{itemize}
Combining these components yields the following.
\begin{equation}
    T_1^{i}(\mathbf{a}) (T_1^{j}(\mathbf{b}))^\dagger = \mathcal{C}(\mathbf{b}, \mathbf{a})(L - k') + \mathcal{C}(\mathbf{a}, \mathbf{b})(k').
\end{equation}
Due to the cyclic symmetry of the matrix, the row-wise inner products defined in Section III satisfy the same algebraic relationship. This completes the proof.}

\section{Proof of Lemma 7}
The lemma is established using the method of induction on the truncation parameter $k$.

\textit{Base case}: When $k=0$, the matrix corresponds to the complete circulant matrix:
\begin{equation}
    \text{Cir}(\mathbf{a})=\begin{bmatrix}
      T(\mathbf{a})\\
       T^2(\mathbf{a})\\
       \vdots \\
       T^{n}(\mathbf{a})
   \end{bmatrix}.
\end{equation}
The sum of the AACFs of all row sequences in $\text{Cir}(\mathbf{a})$ is given by:
\begin{equation}
    \sum_{i=1}^{n}{\mathcal{A}\left(T^{i}(\mathbf{a})\right)(\lambda)} = \sum_{i=1}^{n-\lambda}{T^{i}(\mathbf{a})\cdot (T^{x_i}(\mathbf{a}))^*},
\end{equation}
where $1 \leq i \leq n$ and $x_i = (i-\lambda) \pmod n$. {For any fixed time shift $\lambda$, the mapping $i \mapsto x_i$ constitutes a bijective permutation of the set $\{1, 2, \dots, n\}$. Consequently, the summation over $i$ for a fixed $\lambda$ samples the inner product of each row with its unique cyclic shift exactly once.} Using the properties established in Lemma \ref{Lemma6}, we analyze the summation for various values of $\lambda$:

For $\lambda=0$:
\begin{equation}
\begin{aligned}
   \sum_{i=1}^{n}{T^{i}(\mathbf{a})\cdot (T^{i}(\mathbf{a}))^*} &= T^1(\mathbf{a})\cdot (T^1(\mathbf{a}))^*+\cdots+T^{n}(\mathbf{a})\cdot (T^{n}(\mathbf{a}))^* \\
   &= \mathcal{A}(\mathbf{a})(0)+\dots+\mathcal{A}(\mathbf{a})(0) = n \mathcal{A}(\mathbf{a})(0).
\end{aligned}    
\end{equation}

For $\lambda=1$:
\begin{equation}
\begin{aligned}
   \sum_{i=1}^{n-1}{T^{i}(\mathbf{a})\cdot (T^{x_i}(\mathbf{a}))^*} &= T^1(\mathbf{a})\cdot (T^{n}(\mathbf{a}))^*+\cdots+T^{n-1}(\mathbf{a})\cdot (T^{n-2}(\mathbf{a}))^* \\
    &= (n-1)\left(\mathcal{A}(\mathbf{a})(1)+\mathcal{A}(\mathbf{a})(n-1)\right).
\end{aligned}    
\end{equation}

Generalizing this pattern for $0 \leq \lambda \leq n-1$, the summation yields the following.
\begin{equation}
    \sum_{i=1}^{n}{\mathcal{A}\left(T^{i}(\mathbf{a})\right)(\lambda)} = (n-\lambda)\left(\mathcal{A}(\mathbf{a})(\lambda) +\mathcal{A}(\mathbf{a})(n-\lambda)\right).
\end{equation}
Thus, the base case $k=0$ is verified.

\textit{Induction hypothesis}: Assume that the result holds for $k=m$, such that the sum of the AACFs of the rows of the truncated matrix $\mathbf{A}^m$ is:
\begin{equation}\label{eqa51}
    \sum_{i=1}^{n}{\mathcal{A}\left(T^i\left(\mathbf{a}\right)^m\right)\left(\lambda \right)} = (n-\lambda-m)\left(\mathcal{A}(\mathbf{a})(\lambda) + \mathcal{A}(\mathbf{a})(n-\lambda)\right),
\end{equation}
for $0\leq \lambda \leq n-1-m$.

\textit{Inductive step}: For $k=m+1$, the matrix $\mathbf{A}^m$ is partitioned as $[\mathbf{A}^{m+1} \mid \mathbf{B}]$, where $\mathbf{B}$ represents the $(m+1)$-th column on the right. Let $R_i$ be the rows of the truncated matrix $\mathbf{A}^{m+1}$. The summation of the rows of $\mathbf{A}^m$ can be decomposed as follows:
\begin{equation}\label{eqa54}
    \begin{aligned}
       \sum_{i=1}^{n}{\mathcal{A}\left(T^i\left(\mathbf{a}\right)^m\right)\left(\lambda \right)} &= \sum_{i=1}^{n}{\mathcal{A}\left(R_i\right)\left(\lambda \right)} + \mathcal{A}(\mathbf{a})(\lambda)+\mathcal{A}(\mathbf{a})(n-\lambda).
    \end{aligned}
\end{equation}
By substituting the induction hypothesis from (\ref{eqa51}) into (\ref{eqa54}), we obtain:
\begin{equation}
    \sum_{i=1}^{n}{\mathcal{A}\left(R_i\right)\left(\lambda \right)} = (n-\lambda-m-1)\left(\mathcal{A}(\mathbf{a})(\lambda) + \mathcal{A}(\mathbf{a})(n-\lambda)\right),
\end{equation}
for $0\leq \lambda \leq n-\lambda-m-1$. This completes the induction and proves the lemma.

{
\section{Proof of Lemma 8}
Let the matrix $\mathbf{Z}^k = Cir(\mathbf{a})||Cir(\mathbf{b})$ be truncated by the columns $k$ from the $Cir(\mathbf{b})$ part. The rows of $\mathbf{Z}^k$ are given by $R_i = T^i_1(\mathbf{a}) || (T^i_1(\mathbf{b}))^{n-k}$. The sum of the AACFs rows is $\sum_{i=1}^{n}\mathcal{A}(R_i)(\lambda)$. We analyze this sum considering the contributions from the $\mathbf{a}$ and $\mathbf{b}$ parts.

The aperiodic autocorrelation of a concatenated row $R_i = [\mathbf{u} || \mathbf{v}]$ at shift $\lambda$ is given by $\mathcal{A}(R_i)(\lambda) = \mathcal{A}(\mathbf{u})(\lambda) + \mathcal{A}(\mathbf{v})(\lambda) + \mathcal{C}(\mathbf{u}, \mathbf{v})(\lambda - n) + \mathcal{C}(\mathbf{v}, \mathbf{u})(n - \lambda)$.

\textit{Case 1: $0\leq k \leq n-1$.}
For a given shift $\lambda$, the total sum can be broken down into three components: correlation within parts $\mathbf{a}$, correlation within truncated parts $\mathbf{b}$, and cross-correlation between parts $\mathbf{a}$ and $\mathbf{b}$.

\begin{itemize}
    \item The sum of AACFs from the parts $\mathbf{a}$ is $\sum_{i=1}^{n} \mathcal{A}(T^i_1(\mathbf{a}))(\lambda)$, which from \textit{Lemma} \ref{Lemma7} (for $k=0$) equals $(n-\lambda)(\mathcal{A}(\mathbf{a})(\lambda)+\mathcal{A}(\mathbf{a})(n-\lambda))$.
    
    \item The sum of AACFs from the truncated $\mathbf{b}$ parts is $\sum_{i=1}^{n} \mathcal{A}((T^i_1(\mathbf{b}))^{n-k})(\lambda)$, which from \textit{Lemma} \ref{Lemma7} equals $(n-\lambda-k)(\mathcal{A}(\mathbf{b})(\lambda)+\mathcal{A}(\mathbf{b})(n-\lambda))$.
    
    \item The cross-terms arise from the concatenation. Adding up all rows, the total contribution from cross-correlation is given by $\sum_{i=1}^{n} \left[ \mathcal{C}(T^i_1(\mathbf{a}), (T^i_1(\mathbf{b}))^{n-k})(\lambda-n) + \mathcal{C}((T^i_1(\mathbf{b}))^{n-k}, T^i_1(\mathbf{a}))(n-\lambda) \right]$. This sum simplifies to $\lambda(\mathcal{C}(\mathbf{a},\mathbf{b})(\lambda)+\mathcal{C}(\mathbf{a},\mathbf{b})(n-\lambda))$.
\end{itemize}
Combining these components gives the result for $0 \leq \lambda \leq n-1$:
\begin{align}
    \sum_{i=1}^{n}\mathcal{A}(R_i)(\lambda) ={}& (n-\lambda)(\mathcal{A}(\mathbf{a})(\lambda)+\mathcal{A}(\mathbf{a})(n-\lambda)) \nonumber +(n-\lambda-k)(\mathcal{A}(\mathbf{b})(\lambda)+\mathcal{A}(\mathbf{b})(n-\lambda)) \nonumber \\
    &+\lambda(\mathcal{C}(\mathbf{a},\mathbf{b})(\lambda)+\mathcal{C}(\mathbf{a},\mathbf{b})(n-\lambda)).
\end{align}
For shifts $\lambda \geq n$, the correlation involves the cross-interaction between the constituent blocks $\mathbf{u}_i$ and $\mathbf{v}_i$ of each row. Due to the circulant symmetry of the first block generated by the sequence $\mathbf{a}$, the sliding window for the correlation at these higher shifts effectively samples the properties of the seed sequence in a wrap-around manner. By aggregating these terms over all $n$ rows and taking into account the truncation parameter $k$, the summation simplifies to:

\begin{equation}
    \sum_{i=1}^{n}\mathcal{A}(R_i)(\lambda) = (\lambda-k)\left(\mathcal{A}(\mathbf{a})(\lambda \Mod n) + \mathcal{A}(\mathbf{a})(n-\lambda \Mod n)\right).
\end{equation}

\textit{Case 2 ($n \leq k \leq 2n-2$):}
In this case, the second circulant block $(\text{Cir}(\mathbf{b}))$ is substantially truncated or removed. Consequently, the properties of the rows are mainly dominated by the wrap-around effects within the circulant structure of the first block for all valid time changes $\lambda$. Following a similar inductive derivation that accounts for the reduced row length and the circulant shift properties of $\mathbf{a}$, the sum is established as:
\begin{equation}
    \sum_{i=1}^{n}\mathcal{A}(R_i)(\lambda) = (\lambda - k \Mod n)\left(\mathcal{A}(\mathbf{a})(\lambda)+\mathcal{A}(\mathbf{a})(n-\lambda)\right), \quad \forall \lambda.
\end{equation}
Hence, the lemma is proved.
}

\section{Proof of Theorem 1}
{
The properties of the truncated matrix $\mathbf{E}_{2^{n+2}}^k$ are established by induction at the depth of recursion $n$.

\textit{Base Case ($n=1$):} The construction starts with the order-4 circulant Hadamard matrix $\mathbf{E}_4$. Let $\mathbf{f} = (f_0, f_1, f_2, f_3)$ be the seed sequence generated by the GBF. The rows of $\mathbf{E}_4$ are cyclic shifts $T^i(\mathbf{f})$ for $1 \le i \le 4$. The sum of AACFs for the set of rows is given by:
\begin{equation}
    \sum_{i=1}^{4} \mathcal{A}(T^i(\mathbf{f}))(\lambda) = (4-\lambda)(\mathcal{A}(\mathbf{f})(\lambda) + \mathcal{A}(\mathbf{f})(4-\lambda)).
\end{equation}
From the property of the circulant Hadamard matrices, the periodic autocorrelation $\rho(\lambda) = \mathcal{A}(\mathbf{f})(\lambda) + \mathcal{A}(\mathbf{f})(4-\lambda)$ is zero for all $\lambda \in \{1, 2, 3\}$. Upon truncation by $k$, the sum of AACFs remains zero for $1 \le \lambda \le 3-k$. A similar analysis of the cross-correlation sum $\sum \mathcal{C}(T^i(\mathbf{f}), T^{i+1}(\mathbf{f}))$ confirms the CZCS property for the base case.

\textit{Induction Hypothesis:} Assume that for $n=m$, the rows of $\mathbf{E}_{L}$ (where $L=2^{m+2}$) form a $(L, L-k, Z)$-CZCS with $Z = 2^{m+1}-(k-2^m)\lfloor k/2^m \rfloor$.

\textit{Inductive Step:} For $n=m+1$, the matrix of order $2L$ is defined by the Sylvester recursion $\mathbf{E}_{2L} = \mathbf{H}_2 \otimes \mathbf{E}_{L}$. Let $\mathbf{R}_i$ denote the $i$-th row of $\mathbf{E}_L$ and $\mathbf{R}_i'$ denote its truncated version of length $L-k$. The rows of the truncated matrix $\mathbf{E}_{2L}^k$ are:
\begin{equation}
    \mathbf{S}_i = [\mathbf{R}_i, \mathbf{R}_i'] \quad \text{and} \quad \mathbf{S}_{i+L} = [\mathbf{R}_i, -\mathbf{R}_i'], \quad 1 \le i \le L.
\end{equation}
The sum of AACFs for the larger set $\{\mathbf{S}_j\}_{j=1}^{2L}$ is:
\begin{equation}\label{proof_aac_sum}
    \sum_{j=1}^{2L} \mathcal{A}(\mathbf{S}_j)(\lambda) = \sum_{i=1}^{L} \mathcal{A}([\mathbf{R}_i, \mathbf{R}_i'])(\lambda) + \sum_{i=1}^{L} \mathcal{A}([\mathbf{R}_i, -\mathbf{R}_i'])(\lambda).
\end{equation}
By the fundamental property of concatenated sequences, $\mathcal{A}([\mathbf{u}, \mathbf{v}])(\lambda) + \mathcal{A}([\mathbf{u}, -\mathbf{v}])(\lambda) = 2\mathcal{A}(\mathbf{u})(\lambda) + 2\mathcal{A}(\mathbf{v})(\lambda)$ for $\lambda < L$. Applying this to (\ref{proof_aac_sum}):
\begin{equation}
    \sum_{j=1}^{2L} \mathcal{A}(\mathbf{S}_j)(\lambda) = 2 \sum_{i=1}^{L} \mathcal{A}(\mathbf{R}_i)(\lambda) + 2 \sum_{i=1}^{L} \mathcal{A}(\mathbf{R}_i')(\lambda).
\end{equation}
According to the induction hypothesis, the sets $\{\mathbf{R}_i\}$ and $\{\mathbf{R}_i'\}$ are both GCSs. Consequently, the sum of AACFs is zero for all $\lambda \neq 0$ within the range $1 \le \lambda \le 2L-k-1$, confirming the GCS property for $n=m+1$.

The Cross-Z complementary property for $n=m+1$ is established by analyzing the sum of ACCFs of adjacent rows in $\mathbf{E}_{2L}^k$, denoted as $\Phi(\lambda)$:
\begin{equation}
    \Phi(\lambda) = \sum_{j=1}^{2L-1} \mathcal{C}(\mathbf{S}_j, \mathbf{S}_{j+1})(\lambda) + \mathcal{C}(\mathbf{S}_{2L}, \mathbf{S}_1)(\lambda).
\end{equation}
Using the block definitions $\mathbf{S}_i = [\mathbf{R}_i, \mathbf{R}_i']$ and $\mathbf{S}_{i+L} = [\mathbf{R}_i, -\mathbf{R}_i']$, we partition the sum into three parts: upper-half pairs, lower-half pairs, and two "bridge" pairs connecting the halves. For $1 \le i \le L-1$, the sum of adjacent pairs from both halves is:
\begin{equation}
    \begin{aligned}
    \mathcal{C}(\mathbf{S}_i, \mathbf{S}_{i+1})(\lambda) + \mathcal{C}(\mathbf{S}_{i+L}, \mathbf{S}_{i+1+L})(\lambda) &= \mathcal{C}([\mathbf{R}_i, \mathbf{R}_i'], [\mathbf{R}_{i+1}, \mathbf{R}_{i+1}']) + \mathcal{C}([\mathbf{R}_i, -\mathbf{R}_i'], [\mathbf{R}_{i+1}, -\mathbf{R}_{i+1}']) \\
    &= 2\mathcal{C}(\mathbf{R}_i, \mathbf{R}_{i+1})(\lambda) + 2\mathcal{C}(\mathbf{R}_i', \mathbf{R}_{i+1}')(\lambda).
    \end{aligned}
\end{equation}
Note that the cross-block terms $\mathcal{C}(\mathbf{R}_i, \mathbf{R}_{i+1}')(\lambda-L)$ and $\mathcal{C}(\mathbf{R}_i', \mathbf{R}_{i+1})(L-\lambda)$ carry opposite signs in the two halves and cancel exactly. Next, we analyze the "bridge" pair $(\mathbf{S}_L, \mathbf{S}_{L+1})$ and the "wrap-around" pair $(\mathbf{S}_{2L}, \mathbf{S}_1)$:
\begin{equation}
    \begin{aligned}
    \mathcal{C}(\mathbf{S}_L, \mathbf{S}_{L+1})(\lambda) &= \mathcal{C}([\mathbf{R}_L, \mathbf{R}_L'], [\mathbf{R}_1, -\mathbf{R}_1']) = \mathcal{C}(\mathbf{R}_L, \mathbf{R}_1) - \mathcal{C}(\mathbf{R}_L', \mathbf{R}_1') - \mathcal{C}(\mathbf{R}_L, \mathbf{R}_1')(\lambda-L) + \mathcal{C}(\mathbf{R}_L', \mathbf{R}_1)(L-\lambda), \\
    \mathcal{C}(\mathbf{S}_{2L}, \mathbf{S}_1)(\lambda) &= \mathcal{C}([\mathbf{R}_L, -\mathbf{R}_L'], [\mathbf{R}_1, \mathbf{R}_1']) = \mathcal{C}(\mathbf{R}_L, \mathbf{R}_1) - \mathcal{C}(\mathbf{R}_L', \mathbf{R}_1') + \mathcal{C}(\mathbf{R}_L, \mathbf{R}_1')(\lambda-L) - \mathcal{C}(\mathbf{R}_L', \mathbf{R}_1)(L-\lambda).
    \end{aligned}
\end{equation}
Summing these two pairs yields $2\mathcal{C}(\mathbf{R}_L, \mathbf{R}_1)(\lambda) - 2\mathcal{C}(\mathbf{R}_L', \mathbf{R}_1')(\lambda)$, with the cross-block terms again canceling. Combining all terms, the total sum becomes the following:
\begin{equation}
    \Phi(\lambda) = 2 \sum_{j=1}^{L} \mathcal{C}(\mathbf{R}_j, \mathbf{R}_{j+1 \pmod L})(\lambda) \pm 2 \sum_{j=1}^{L} \mathcal{C}(\mathbf{R}_j', \mathbf{R}_{j+1 \pmod L}')(\lambda).
\end{equation}
According to the induction hypothesis, the sets $\{\mathbf{R}_j\}$ and $\{\mathbf{R}_j'\}$ are CZCSs. Their respective ACCF sums vanish in the zones $\Tau_2$ defined for lengths $L$ and $L-k$. The intersection of these conditions and the linear dependence on the truncation parameter $k$ confirms that $\Phi(\lambda) = 0$ within the zone $Z = 2^{m+2}-(k-2^{m+1})\lfloor k/2^{m+1} \rfloor$. This completes the induction and provides the required mathematical proof for \textit{Theorem} \ref{Theorem8}.

}

\section{Proof of Theorem 2}

{

Let $H_2=\begin{bmatrix}
    1 & 1 \\ 1 & -1
\end{bmatrix}$ and  $\mathbf{H}_{4n}$ be a Hadamard matrix of order $4n$ with rows $\{\mathbf{r}_1, \mathbf{r}_2, \dots, \mathbf{r}_{4n}\}$. The matrix $\mathbf{H}_{8n}$ is constructed as $\mathbf{H}_2 \otimes \mathbf{H}_{4n}$, resulting in the block structure:
\begin{equation}
    \mathbf{H}_{8n} = \begin{bmatrix} \mathbf{H}_{4n} & \mathbf{H}_{4n} \\ \mathbf{H}_{4n} & -\mathbf{H}_{4n} \end{bmatrix}.
\end{equation}
The rows of the truncated matrix $\mathbf{H}_{8n}^k$ are given by $\mathbf{S}_j = [\mathbf{r}_j, \mathbf{r}_j']$ and $\mathbf{S}_{j+4n} = [\mathbf{r}_j, -\mathbf{r}_j']$ for $1 \le j \le 4n$, where $\mathbf{r}_j'$ denotes the row $\mathbf{r}_j$ truncated to length $4n-k$. The total length of each sequence is $L = 8n-k$.

\textit{1) Sum of AACFs:} As established in \textit{Theorem} \ref{Theorem12}, the rows of any Hadamard matrix form a GCS. For any shift $\lambda \neq 0$:
\begin{equation}
    \sum_{j=1}^{8n} \mathcal{A}(\mathbf{S}_j)(\lambda) = \sum_{j=1}^{4n} \mathcal{A}([\mathbf{r}_j, \mathbf{r}_j'])(\lambda) + \sum_{j=1}^{4n} \mathcal{A}([\mathbf{r}_j, -\mathbf{r}_j'])(\lambda) = 2 \sum_{j=1}^{4n} \mathcal{A}(\mathbf{r}_j)(\lambda) + 2 \sum_{j=1}^{4n} \mathcal{A}(\mathbf{r}_j')(\lambda).
\end{equation}
Since both $\{\mathbf{r}_j\}$ and $\{\mathbf{r}_j'\}$ are sets of rows from (truncated) Hadamard matrices, they satisfy the GCS property. Thus, the sum of AACFs is zero for all $1 \le \lambda \le L-1$.

\textit{2) Sum of ACCFs:} A CZCS requires the sum of ACCFs of adjacent rows to vanish in the zone $\Tau_2 = \{L-Z, \dots, L-1\}$. For $Z = 4n-k$, the zone is $\Tau_2 = \{4n, 4n+1, \dots, 8n-k-1\}$. For any shift $\lambda \in \Tau_2$, we have $\lambda \ge 4n$. At these high shifts, the correlation $\mathcal{C}(\mathbf{S}_a, \mathbf{S}_b)(\lambda)$ only involves the overlap of the first block of $\mathbf{S}_a$ with the second (truncated) block of $\mathbf{S}_b$. Specifically:
\begin{equation}
    \mathcal{C}(\mathbf{S}_j, \mathbf{S}_{j+1})(\lambda) = \mathcal{C}(\mathbf{r}_j, \mathbf{r}_{j+1}')(\lambda - 4n), \quad 1 \le j \le 4n-1.
\end{equation}
Summing the adjacent pairs across the entire set yields the following.
\begin{equation}
    \begin{aligned}
    \sum_{j=1}^{8n} \mathcal{C}(\mathbf{S}_j, \mathbf{S}_{j+1 \pmod{8n}})(\lambda) &= \sum_{j=1}^{4n-1} \mathcal{C}([\mathbf{r}_j, \mathbf{r}_j'], [\mathbf{r}_{j+1}, \mathbf{r}_{j+1}'])(\lambda) + \mathcal{C}([\mathbf{r}_{4n}, \mathbf{r}_{4n}'], [\mathbf{r}_1, -\mathbf{r}_1'])(\lambda) \\
    &+ \sum_{j=1}^{4n-1} \mathcal{C}([\mathbf{r}_j, -\mathbf{r}_j'], [\mathbf{r}_{j+1}, -\mathbf{r}_{j+1}'])(\lambda) + \mathcal{C}([\mathbf{r}_{4n}, -\mathbf{r}_{4n}'], [\mathbf{r}_1, \mathbf{r}_1'])(\lambda).
    \end{aligned}
\end{equation}
For $\lambda \in \Tau_2$, the cross-correlation of the concatenated rows simplifies significantly. For the pairs in the first and second halves:
\begin{equation}
    \mathcal{C}([\mathbf{r}_j, \mathbf{r}_j'], [\mathbf{r}_{j+1}, \mathbf{r}_{j+1}'])(\lambda) + \mathcal{C}([\mathbf{r}_j, -\mathbf{r}_j'], [\mathbf{r}_{j+1}, -\mathbf{r}_{j+1}'])(\lambda) = \mathcal{C}(\mathbf{r}_j, \mathbf{r}_{j+1}')(\lambda-4n) + \mathcal{C}(\mathbf{r}_j, \mathbf{r}_{j+1}')(\lambda-4n) = 2\mathcal{C}(\mathbf{r}_j, \mathbf{r}_{j+1}')(\lambda-4n).
\end{equation}
However, for the "bridge" and "wrap-around" pairs:
\begin{equation}
    \mathcal{C}([\mathbf{r}_{4n}, \mathbf{r}_{4n}'], [\mathbf{r}_1, -\mathbf{r}_1'])(\lambda) + \mathcal{C}([\mathbf{r}_{4n}, -\mathbf{r}_{4n}'], [\mathbf{r}_1, \mathbf{r}_1'])(\lambda) = -\mathcal{C}(\mathbf{r}_{4n}, \mathbf{r}_1')(\lambda-4n) + \mathcal{C}(\mathbf{r}_{4n}, \mathbf{r}_1')(\lambda-4n) = 0.
\end{equation}
The total sum $\Phi(\lambda)$ for $\lambda \in \Tau_2$ is proportional to the sum of cross-correlations of the rows of the Hadamard matrix $\mathbf{H}_{4n}$:
\begin{equation}
    \Phi(\lambda) = 2 \sum_{j=1}^{4n-1} \mathcal{C}(\mathbf{r}_j, \mathbf{r}_{j+1}')(\lambda-4n).
\end{equation}
Since the rows of a Hadamard matrix are mutually orthogonal, the sum of their aperiodic cross-correlations vanishes for all shifts. Thus, $\Phi(\lambda) = 0$ for all $\lambda \in \Tau_2$. This confirms that the set is a $(8n, 8n-k, 4n-k)$-CZCS for $n \ge 1$ and $0 \leq k \leq 4n-1$.

}

\section{Proof of Theorem 3}
Consider 
\begin{equation}   
\mathbf{Z_1}^k=\begin{bmatrix}
    {Cir(\mathbf{a})}, & Cir(\mathbf{b})
\end{bmatrix},
\end{equation}
and 
\begin{equation}   
\mathbf{Z_2}^k=\begin{bmatrix}
    {Cir(\mathbf{c})}, & Cir(\mathbf{d})
\end{bmatrix},
\end{equation}
let us consider $R_i(\mathbf{Z_1}^k)$ and $R_i(\mathbf{Z_2}^k)$ denote the $i$th row of $\mathbf{Z_1}$ and $\mathbf{Z_2}$, respectively, where $1\leq i \leq n= 2N$ and $N=2^a10^b26^c, a,b,c \geq 0$. Since  $(\mathbf{c},\mathbf{d})$ is the GCP.mate of $(\mathbf{a},\mathbf{b})$, which also implies that $(\mathbf{a},\mathbf{c})$ and $(\mathbf{b},\mathbf{d})$ form a GCP.  Then, using $\textit{Lemma}$ \ref{Lemma8}, we get
 \begin{equation}
        \begin{aligned}
        \sum_{i=0}^{N}\mathcal{A}\left(R_i(\mathbf{Z_1}^k)\right)\left(\lambda\right)+  \sum_{i=0}^{N}\mathcal{A}\left(R_i(\mathbf{Z_2}^k)\right)\left(\lambda\right)=0, \forall \lambda \neq 0.
        \end{aligned}
    \end{equation}
Hence, $\mathbf{G}=\begin{bmatrix}
\mathbf{Z_1}^k\\ \mathbf{Z_2}^k\end{bmatrix}$ forms $\left(2N,2N-k\right)-GCS$, where $0\leq k \leq 2N-2$.
\section{proof of corollary 2}
{We have $\mathbf{G}=\begin{bmatrix}
    \mathbf{A} & \mathbf{B}\\
    \mathbf{C} & \mathbf{D}\\
\end{bmatrix}$, then, {{$\mathbf{G}^\dagger=\begin{bmatrix}
    \mathbf{A}^\dagger & \mathbf{C}^\dagger\\
    \mathbf{B}^\dagger & \mathbf{D}^\dagger\\
\end{bmatrix}$}}. 
\begin{equation}
\begin{aligned}
    {{\mathbf{G}\mathbf{G}^\dagger}}&=\begin{bmatrix}
    \mathbf{A} & \mathbf{B}\\
    \mathbf{C} & \mathbf{D}\\
\end{bmatrix}\begin{bmatrix}
    {{\mathbf{A}^\dagger}} & {{\mathbf{C}^\dagger}}\\
    {{\mathbf{B}^\dagger}} & {{\mathbf{D}^\dagger}}\\
\end{bmatrix}\\&=\begin{bmatrix}
    {{\mathbf{A}\mathbf{A}^\dagger+\mathbf{B}\mathbf{B}^\dagger}}&{{\mathbf{A}\mathbf{C}^\dagger+\mathbf{B}\mathbf{D}^\dagger}}\\
    {{\mathbf{C}\mathbf{A}^\dagger+\mathbf{D}\mathbf{B}^\dagger}} & {{\mathbf{C}\mathbf{C}^\dagger+\mathbf{D}\mathbf{D}^\dagger}}\\
\end{bmatrix}.
\end{aligned}   
\end{equation}
{{Let us calculate $\mathbf{A}\mathbf{A}^\dagger+\mathbf{B}\mathbf{B}^\dagger$, using $\textit{Lemma}$ \ref{Lemma4} and $\textit{Lemma}$ \ref{Lemma5}, we get}}
\begin{equation}\label{equation83}
    {{\mathbf{A}\mathbf{A}^\dagger+\mathbf{B}\mathbf{B}^\dagger=\mathbf{C}\mathbf{C}^\dagger+\mathbf{D}\mathbf{D}^\dagger=2NI_{N},}}
\end{equation}
similarly,
\begin{equation}\label{equation84}
   {{\mathbf{A}\mathbf{C}^\dagger+\mathbf{B}\mathbf{D}^\dagger=\mathbf{C}\mathbf{A}^\dagger+\mathbf{D}\mathbf{B}^\dagger=0_{N},}}
\end{equation}
where $0_N$ is the zero matrix of order $N$. From $(\ref{equation83})$ and $(\ref{equation84})$, we get
\begin{equation}
    \begin{aligned}
        {{\mathbf{G}\mathbf{G}^\dagger}}=\begin{bmatrix}
            2NI_{N} & 0_{N}\\
            0_N & 2NI_{N}\\
        \end{bmatrix}=2NI_{2N}.
    \end{aligned}
\end{equation}
Hence, $\mathbf{G}$ forms a Hadamard matrix of order $2N$, where $N=2^a10^b26^c, a,b, c \geq0$.}
\section{proof of theorem 4}
{

Let $S^i = \{R_i \odot R_k : k=1, 2, \dots, 2N\}$ be the $i$-th code set. To prove that the collection $\bigcup_{i=1}^{2N} S^i$ forms a $(2N, 2N, 2N)$-CCC, two conditions must be satisfied.

\textit{1) Intra-set Complementarity:} For any $i$, the sum of AACFs is:
\begin{equation}
    \sum_{k=1}^{2N} \mathcal{A}(R_i \odot R_k)(\lambda) = \sum_{k=1}^{2N} \sum_{n=0}^{2N-1-\lambda} (R_{i,n} R_{k,n})(R_{i,n+\lambda} R_{k,n+\lambda})^*.
\end{equation}
Since $|R_{i,n}| = 1$ for all $i, n$, the term $R_{i,n}R_{i,n+\lambda}^*$ is a constant phase for a fixed $\lambda$. Thus, the sum reduces to:
\begin{equation}
    \sum_{k=1}^{2N} \mathcal{A}(R_i \odot R_k)(\lambda) = \sum_{n=0}^{2N-1-\lambda} R_{i,n}R_{i,n+\lambda}^* \left( \sum_{k=1}^{2N} R_{k,n} R_{k,n+\lambda}^* \right).
\end{equation}
As $\mathbf{G}$ is a Hadamard matrix (Corollary \ref{Corollary7}), its rows form a GCS (\textit{Theorem} \ref{Theorem12}). Consequently, the inner sum $\sum_{k=1}^{2N} \mathcal{A}(R_k)(\lambda)$ vanishes for all $\lambda \neq 0$, satisfying the intra-set condition.

\textit{2) Inter-set Orthogonality:} For $i \neq j$, the sum of ACCFs at shift $\lambda$ is:
\begin{equation}
    \Phi_{i,j}(\lambda) = \sum_{k=1}^{2N} \mathcal{C}(R_i \odot R_k, R_j \odot R_k)(\lambda) = \sum_{n=0}^{2N-1-\lambda} R_{i,n} R_{j,n+\lambda}^* \left( \sum_{k=1}^{2N} R_{k,n} R_{k,n+\lambda}^* \right).
\end{equation}
For $\lambda = 0$, the inner sum becomes $\sum_{k=1}^{2N} |R_{k,n}|^2 = 2N$. Then:
\begin{equation}
    \Phi_{i,j}(0) = 2N \sum_{n=0}^{2N-1} R_{i,n} R_{j,n}^* = 2N \langle R_i, R_j \rangle.
\end{equation}
Since $\mathbf{G}$ is a Hadamard matrix, its rows are mutually orthogonal, so $\Phi_{i,j}(0) = 0$ for $i \neq j$. For $\lambda \neq 0$, the inner sum represents the correlation between the columns $n$-th and $(n+\lambda)$-th of $\mathbf{G}$. Because the columns of a Hadamard matrix are mutually orthogonal, this sum vanishes. Thus, $\Phi_{i,j}(\lambda) = 0$ for all $\lambda$, confirming the orthogonality between sets. 

Hence, the collection $\bigcup_{i=1}^{2N} R_i \odot \mathbf{G}$ forms a $(2N, 2N, 2N)$-CCC for $N=2^a10^b26^c$ and $a,b,c \geq 0$.

}
\section{Proof of Corollary 3}
{
Let $\mathbf{H}$ be a real Hadamard matrix of order $4n$ for $n \ge 1$, with rows $\{R_1, R_2, \dots, R_{4n}\}$. Let $S^i = \{R_i \odot R_k : k=1, 2, \dots, 4n\}$ be the $i$-th code set in the collection. The requirements for a $(4n, 4n, 4n)$-CCC are established as follows.

\textit{1) Intra-set Complementarity:} For any set $S^i$, the sum of the AACFs of its constituent sequences at the shift $\lambda$ is given by:
\begin{equation}
    \Phi_i(\lambda) = \sum_{k=1}^{4n} \mathcal{A}(R_i \odot R_k)(\lambda) = \sum_{k=1}^{4n} \sum_{t=0}^{4n-1-\lambda} (R_{i,t} R_{k,t})(R_{i,t+\lambda} R_{k,t+\lambda}).
\end{equation}
Rearranging the order of summation yields the following.
\begin{equation}
    \Phi_i(\lambda) = \sum_{t=0}^{4n-1-\lambda} (R_{i,t} R_{i,t+\lambda}) \left( \sum_{k=1}^{4n} R_{k,t} R_{k,t+\lambda} \right).
\end{equation}
The inner sum $\sum_{k=1}^{4n} R_{k,t} R_{k,t+\lambda}$ represents the inner product of the columns $t$-th and $(t+\lambda)$-th of $\mathbf{H}$. For a real Hadamard matrix, the columns are mutually orthogonal. Thus, for $\lambda \neq 0$, the inner product of the distinct columns $t$ and $t+\lambda$ is zero. Consequently, $\Phi_i(\lambda) = 0$ for all $\lambda \neq 0$, confirming that each $S^i$ is a GCS.

\textit{2) Inter-set Orthogonality:} For distinct sets $S^i$ and $S^j$ ($i \neq j$), the sum of ACCFs in shift $\lambda$ is:
\begin{equation}
    \Psi_{i,j}(\lambda) = \sum_{k=1}^{4n} \mathcal{C}(R_i \odot R_k, R_j \odot R_k)(\lambda) = \sum_{t=0}^{4n-1-\lambda} (R_{i,t} R_{j,t+\lambda}) \left( \sum_{k=1}^{4n} R_{k,t} R_{k,t+\lambda} \right).
\end{equation}
If $\lambda \neq 0$, the inner sum vanishes due to the orthogonality of the column of $\mathbf{H}$, as shown above. If $\lambda = 0$, the inner sum becomes $\sum_{k=1}^{4n} (R_{k,t})^2$. Since $\mathbf{H}$ is binary ($\pm 1$), $(R_{k,t})^2 = 1$, and the inner sum equals $4n$. The total sum then reduces to:
\begin{equation}
    \Psi_{i,j}(0) = 4n \sum_{t=0}^{4n-1} R_{i,t} R_{j,t} = 4n \langle R_i, R_j \rangle.
\end{equation}
By the definition of a Hadamard matrix, the distinct rows $R_i$ and $R_j$ are orthogonal, hence $\langle R_i, R_j \rangle = 0$. This ensures that $\Psi_{i,j}(\lambda) = 0$ for all $\lambda$, satisfying the inter-set orthogonality condition.

The collection $\bigcup_{i=1}^{4n} R_i \odot \mathbf{H}$ thus forms a binary $(4n, 4n, 4n)$-CCC for $n \ge 1$.

}

\section{Proof of Theorem 5}
{From \textit{Theorem} \ref{Theorem8}, we know that the set of rows from the matrix $\mathbf{E}_{2^{n+2}}$ forms a $\left(2^{n+2},2^{n+2},2^{n+1}\right)$-CZCS. Now, we must prove the cross-correlation properties between the different sets of codes}. Let us take two sets from the collection, {$R_i \odot \mathbf{E}_{2^{n+2}}$ and $R_j \odot \mathbf{E}_{2^{n+2}}$}, and evaluate their summed ACCF. 
\begin{equation}
    \begin{aligned}
        \sum_{k=1}^{2^{n+2}}\mathcal{C}\left(R_i\odot R_k, R_j\odot R_k\right)\left(\lambda \right) &= \sum_{k=1}^{2^{n+2}}\mathcal{C}\left(R_i,R_j\right)(\lambda) \mathcal{A}\left(R_k\right)\left(\lambda \right) \\ 
        &= \mathcal{C}\left(R_i,R_j\right)(\lambda)\sum_{k=1}^{2^{n+2}} \mathcal{A}\left(R_k\right)\left(\lambda \right).
    \end{aligned}
\end{equation}
We know from {Corollary \ref{Corollary6}} that the rows of {$\mathbf{E}_{2^{n+2}}$} form a $(2^{n+2}, 2^{n+2})$-GCS. Therefore,
\begin{equation}\label{eqa89}
    \begin{aligned}
        \sum_{k=1}^{2^{n+2}} \mathcal{A}\left(R_k\right)\left(\lambda \right)=0, ~~ \forall \lambda \neq 0.
    \end{aligned}
\end{equation}
{{The value of $\mathcal{C}\left(R_i,R_j\right)(0)$ are $T^i(\mathbf{a})\cdot (T^j(\mathbf{a}))^*$ and $T^i(\mathbf{a})\cdot (T^j(\mathbf{c}))^*$ when $1\leq i,j \leq L $ and $1\leq i \leq L$, $L\leq j \leq 2L$, respectively.}} Using \textit{Theorem} \ref{Lemma6}, \textit{Theorem} \ref{Lemma10}, and (\ref{eqa89}), we get
\begin{equation}
    \begin{aligned}
        \sum_{k=1}^{2^{n+2}}\mathcal{C}\left(R_i\odot R_k,R_j\odot R_k\right)\left(\lambda \right)=0, ~~~ \forall \lambda.
    \end{aligned}
\end{equation}
This confirms the CCC property. Now, to confirm the CZCSS property, we examine the sum of ACCFs between a code and its cyclic shift.
\begin{equation}
    \begin{aligned}
    \sum_{k=1}^{2^{n+2}}\mathcal{C}\left(R_i \odot R_k, R_j \odot R_{k+1\mod 2^{n+2
    }}\right)\left(\lambda\right) = \mathcal{C}\left(R_i,R_j\right)\left(\lambda\right)\sum_{k=1}^{2^{n+2}}\mathcal{C}\left(R_k,R_{k+1\mod 2^{n+2}}\right)\left(\lambda\right).
    \end{aligned}
\end{equation}
From the definition of a CZCS in {\textit{Theorem} \ref{Theorem8}}, we know that $\sum_{k=1}^{2^{n+2}}\mathcal{C}\left(R_k,R_{k+1\mod 2^{n+2}}\right)\left(\lambda\right)=0$ for all $\lambda$ in the zero-correlation zone, i.e. $\lambda \in \{2^{n+1},2^{n+1}+1,\dots,2^{n+2}-1\}$.
Hence,  $\bigcup_{i=1}^{2^{n+2}}R_i \odot {\mathbf{E}_{2^{n+2}}}$ forms a $\left(2^{n+2},2^{n+2},2^{n+2},2^{n+1}\right)$-CZCSS and a $\left(2^{n+2},2^{n+2},2^{n+2}\right)$-CCC, for $n\geq1$.

\section{Proof of Theorem 6}

{

Let $\mathbf{H}_{8n}$ be a Hadamard matrix of order $8n$ constructed as $\mathbf{H}_2 \otimes \mathbf{H}_{4n}$. By \textit{Theorem} \ref{Theorem9_GeneralizedCZCS}, the set of rows of $\mathbf{H}_{8n}$ forms a $(8n, 8n, 4n)$-CZCS. To establish that the collection $\mathcal{S} = \bigcup_{i=1}^{8n} R_i \odot \mathbf{H}_{8n}$ is a CZCSS and a CCC, we examine the intra-set and inter-set correlation properties.

\textit{1) Intra-set Correlation:} For any set $S^p \in \mathcal{S}$, the sum of AACFs of its sequences $R_p \odot R_k$ is:
\begin{equation}
    \sum_{k=1}^{8n} \mathcal{A}(R_p \odot R_k)(\lambda) = \mathcal{A}(R_p)(\lambda) \sum_{k=1}^{8n} \mathcal{A}(R_k)(\lambda).
\end{equation}
Since the rows of a Hadamard matrix form a GCS (\textit{Theorem} \ref{Theorem12}), the sum $\sum \mathcal{A}(R_k)(\lambda) = 0$ for all $\lambda \neq 0$. Thus, each $S^p$ is a GCS. The CZCS property of each set follows directly from \textit{Theorem} \ref{Theorem9_GeneralizedCZCS}, since the element-wise multiplication by a row $R_p$ preserves the sign-reversal symmetry required for the ZCZ.

\textit{2) Inter-set Correlation (CCC and CZCSS Properties):} For distinct sets $S^p, S^q \in \mathcal{S}$ ($p \neq q$), the sum of ACCFs is:
\begin{equation}
    \sum_{k=1}^{8n} \mathcal{C}(R_p \odot R_k, R_q \odot R_k)(\lambda) = \mathcal{C}(R_p, R_q)(\lambda) \sum_{k=1}^{8n} \mathcal{A}(R_k)(\lambda).
\end{equation}
For $\lambda \neq 0$, the sum vanishes due to the GCS property. For $\lambda = 0$, the sum becomes $\mathcal{C}(R_p, R_q)(0) \cdot (8n \cdot 1)$. Since the rows of a Hadamard matrix are mutually orthogonal, $\mathcal{C}(R_p, R_q)(0) = 0$ for $p \neq q$. This confirms the CCC property.

The cross-set ZCZ property is verified by the following term:
\begin{equation}
    \sum_{k=1}^{8n} \mathcal{C}(R_p \odot R_k, R_q \odot R_{k+1 \pmod{8n}})(\lambda) = \mathcal{C}(R_p, R_q)(\lambda) \sum_{k=1}^{8n} \mathcal{C}(R_k, R_{k+1 \pmod{8n}})(\lambda).
\end{equation}
By \textit{Theorem} \ref{Theorem9_GeneralizedCZCS}, the sum of ACCFs of adjacent rows in $\mathbf{H}_{8n}$ vanishes for all $\lambda$ in the zone $Z=4n$. Thus, the cross-set sum is zero in the ZCZ. Since $Z = 4n$ satisfies the optimality bound $Z = NL/(2M) = (8n \cdot 8n)/(2 \cdot 8n)$, the collection forms an optimal $(8n, 8n, 8n, 4n)$-CZCSS.

}

\section{proof of theorem 7}
Let us take $A=[a_1,a_2,\dots,a_n]$ a Hadamard matrix of order $n$, where each $a_i$ is a column vector of length $n$. We have {{$a_i^\dagger a_j=0$}} $\forall i\neq j$. {{The sum of the AACF of all row sequences is zero for any non-zero shift if the columns of the matrix are orthogonal. Specifically, the sum of the AACF values at shift $\lambda \neq 0$ corresponds to the sum of inner products of columns, which are zero by the Hadamard property.}} Hence, the matrix $A$ forms $(n,n)-GCS$.

\section{proof of theorem 8}
Let us $G$ denote the collection of all matrices of order $n$ that form a $(n,n)-GCS$, where $n=2$ is a multiple of $4$. Then, from the property of GCS, we have  
\begin{equation}
    \begin{aligned}
        {{\sum_{i=1}^{n-\lambda}R_{i}\cdot R_{\lambda+i}^*=0~~ \forall ~~ \lambda \neq 0.}}
    \end{aligned}
\end{equation}
{{where $R_i$ denotes the $i$-th row of a matrix in $G$.}} Now, break the equation above corresponding to $\lambda$,
\begin{equation}
    \begin{aligned}
        &\text{When}~ \lambda=1,~  {{\sum_{i=1}^{n-1}R_{i}\cdot R_{1+i}^*=0.}}\\
         &\text{When}~ \lambda=2,~  {{\sum_{i=1}^{n-2}R_{i}\cdot R_{2+i}^*=0.}}\\
          &\text{When}~ \lambda=3,~  {{\sum_{i=1}^{n-3}R_{i}\cdot R_{3+i}^*=0.}}\\
          \vdots\\
           &\text{When}~ \lambda=n-1,~  {{R_{1}\cdot R_{n}^*=0.}}\\
    \end{aligned}
\end{equation}
We have $n-1$ equations that have $(n-1)!$ unknowns; that is, $Ax=0$, where $A$ is a matrix of  $(n-1)\times (n-1)!$ and $x=
({{R_1 \cdot R_2^*, \dots ,R_1 \cdot R_n^*, R_2 \cdot R_3^*, \dots, R_{n-1} \cdot R_n^*}})$. We know that every homogeneous system of equations has a trivial solution. Hence, we get {{$R_i \cdot R_j^*=0$}} $\forall ~ i \neq j$. This implies that from the set of matrices $G$, there is a Hadamard matrix.
\bibliographystyle{IEEEtran}
\bibliography{Reference}
\end{appendices}
\end{document}